\documentclass[paper,11pt]{JHEP}
\pdfoutput=1
\usepackage{graphicx}
\usepackage[centertags]{amsmath}
\usepackage{amsfonts}
\usepackage{amssymb} 
\usepackage{amsthm}
\usepackage{slashed}
\usepackage{enumerate}
\usepackage{amsbsy}

\def\one{{\rm 1\kern -.9mm l}}                             %

\def\beq{\begin{equation}}
\def\eeq{\end{equation}}
\def\beqa{\begin{eqnarray}}
\def\eeqa{\end{eqnarray}}
\newcommand{\eqa}{\begin{eqnarray}}
\newcommand{\ena}{\end{eqnarray}}
\newcommand{\p}{\partial}

\newcommand{\R}{{\mathbb{R}}}

\newcommand\blank[1]{}

\newcommand{\bP}{ {\bf P } }
\newcommand{\bh}{ {\bf h } }
\newcommand{\wT}{ \mathbb{T} }
\newcommand{\hY}{ Y }

\newcommand{\eps}{\varepsilon}

\newcommand\ZZ{{\mathbb Z}}

\newcommand{\balpha}{\alpha\kern -6.7pt\alpha}
\newcommand{\bbalpha}{\alpha\kern -4.95pt\alpha}

\newcommand\en{\end{equation}}
\newcommand\bea{\begin{eqnarray}}
\newcommand\eea{\end{eqnarray}}
\newcommand\nn{\nonumber}

\newcommand{\One}{{\hbox{{\rm 1{\hbox to 1.5pt{\hss\rm1}}}}}}
\renewcommand{\One}{{\mathbb 1}}
\renewcommand{\One}{{\rm 1\!\!1}}

\newcommand{\RRe}{\text{Re}}
\newcommand{\IIm}{\text{Im}}
\newcommand{\ba}{\begin{eqnarray}}
\newcommand{\ea}{\end{eqnarray}}

\newcommand{\be}{\begin{equation}}
\newcommand{\ee}{\end{equation}}

\newcommand{\bu}{{\bf u}}

\newcommand{\mT}{{{\mathcal T }}}
\newcommand{\ra}{\textsf{\tiny{H}}}%
\newcommand{\ua}{\textsf{\tiny{V}}}%

\newcommand{\hmu}{\hat{\mu}}
\newcommand{\hB}{\hat{B}}
\newcommand{\bF}{{\bf F }}
\newcommand{\f}{\mathcal{Q}}
\newcommand{\mG}{\mathcal{G}}
\newcommand{\tP}{{\widetilde \bP }}
\renewcommand{\log}{\ln}
\newcommand{\brho}{\rho}
\newcommand{\mQ}{\mathcal{Q}}
\newcommand{\lra}{\leftrightarrow}
\newcommand{\sq}{\phi}
\newcommand{\CG}{{\mathcal G}}
\newcommand{\slG }{{\slashed \CG }}
\newcommand{\TPsi}{{\Psi}}
\renewcommand{\p}{{\boldsymbol{+}}}
\newcommand{\m}{{ \boldsymbol{-}}}
\newcommand{\bpm}{{\boldsymbol{\pm}}}
\newcommand{\bmp}{{\boldsymbol{\mp}}}
\newcommand{\bigra}{\textsf{H}}%
\newcommand{\bigua}{\textsf{V}}%
\newcommand{\arcsinh}{\text{arcsinh}}
\newcommand{\disc}{\text{disc}}
\newcommand{\x}{{ \bf x }}
\newcommand{\thh}{ { \bf k } }
\newcommand{ \beps}{ \boldsymbol{\eps } }

\newcommand\hbeta{{\hat \beta}}
\newcommand\halpha{{\hat  \alpha }}

\newcommand{\bq}{{ \bf p }}
\newcommand{\bG}{{ \bf f }}

\newcommand{\NT}{ { \bar N } }
\newcommand{\InR}{\int_\R}

\def\Xint#1{\mathchoice
{\XXint\displaystyle\textstyle{#1}}%
{\XXint\textstyle\scriptstyle{#1}}%
{\XXint\scriptstyle\scriptscriptstyle{#1}}%
{\XXint\scriptscriptstyle\scriptscriptstyle{#1}}%
\!\int}
\def\XXint#1#2#3{{\setbox0=\hbox{$#1{#2#3}{\int}$ }
\vcenter{\hbox{$#2#3$ }}\kern-.6\wd0}}

\def\dashint{\Xint-}

\title{A Riemann-Hilbert formulation for the finite temperature Hubbard model}
\author{ \centerline{Andrea Cavagli\`a$^{1}$, Martina Cornagliotto$^{1,2}$,
 Massimo Mattelliano$^{1}$ and Roberto Tateo$^{1}$} \\
\centerline{${}^{1}$\sl\small Dipartimento\ di Fisica
and INFN, Universit\`a di Torino, Via P.\
Giuria 1, 10125 Torino, Italy} 
\centerline{${}^{2}$DESY Hamburg, Theory Group, Notkestrasse 85, D-22607 Hamburg, Germany}
\vspace{0.3cm}
\centerline{ \it cavaglia/mattelli/tateo@to.infn.it, martina.cornagliotto@desy.de}
}
\abstract{
Inspired by recent results in the context of AdS/CFT integrability, we reconsider the Thermodynamic Bethe Ansatz 
equations describing the 1D fermionic Hubbard model at finite temperature. We prove that the infinite set of TBA equations are equivalent to a 
simple nonlinear Riemann-Hilbert problem for a finite number of unknown functions. The latter can be transformed into a set of three coupled nonlinear integral equations defined over a finite support, which can be easily solved numerically. We discuss the emergence of an exact Bethe Ansatz and the link between the TBA approach and the results by J\"uttner, Kl\"umper and Suzuki based on the Quantum Transfer Matrix method. We also comment on the analytic continuation mechanism leading to excited states and on the mirror equations describing the finite-size Hubbard model with twisted boundary conditions. 
}

\begin{document}

\section{Introduction}
\label{sec:intro}
The Hubbard model~\cite{Hubbard} arises as an approximate description of correlated electrons in narrow-band materials. Despite its simplicity, it is believed to capture important nonperturbative features of real many-body fermionic systems.
In one dimension, the model is exactly solvable and its relevance for the study of strongly correlated electrons is perhaps comparable to that of the Ising model for magnetism.
 Quite unexpectedly, the Hubbard model has also attracted the attention of high energy physicists due to its multiple connections 
 \cite{StaudacherHubbard, MinahanHubbard, BES, FioravantiHubbard, BeisertHubbard, ArutyunovString, VieiraPenedones} 
 with the integrable spin chains emerging in the context of $\mathcal{N}{=}4$ Super Yang-Mills theory~\cite{MZ,AdSCFTReview,BS}. 
In particular, the exact Bethe Ansatz equations of Lieb and Wu~\cite{LiebWu}, and the corresponding string hypothesis~\cite{Takahashi}, 
have many features in common with the asymptotic Bethe Ansatz of Beisert and Staudacher for the anomalous dimensions of 
single-trace operators with large quantum numbers~\cite{BS}. 

 We start from the 1D Hubbard Hamiltonian written in the form
\begin{equation}
\label{H}
H_0 =
-\, \sum_{i=1}^L \sum_{\sigma=\uparrow,\downarrow}
\left(c^\dagger_{i,\sigma} c_{i+1,\sigma}+
c^\dagger_{i+1,\sigma} c_{i,\sigma}\right)+
 \bu \,  \sum_{i=1}^L (1- 2\,
c^\dagger_{i,\uparrow}c_{i,\uparrow})(1-2\,c^\dagger_{i,\downarrow} c_{i,\downarrow}),
\end{equation}
where $L$ is the length of the chain, $c^\dagger_{i,\sigma}$ and $c_{i,\sigma}$ are fermionic creation-annihilation operators satisfying
\bea
\{c_{i,\sigma},c_{j,\tau}\}=
\{c^\dagger_{i,\sigma},c^\dagger_{j,\tau}\}=0, \,\,\,\,
\{c_{i,\sigma},c^\dagger_{j,\tau}\}=
\delta_{i j}\,\delta_{\sigma \tau},
\eea
with periodic boundary conditions $c_{1, \sigma}=c_{L+1, \sigma}$, $c^{\dagger}_{1, \sigma}=c^{\dagger}_{L+1, \sigma}$, and 
$\bu \ge  0$ is a dimensionless coupling constant proportional to the electric charge. 
For $L$ even, the Hamiltonian (\ref{H}) is $SO(4)$-symmetric \cite{HubbardSO4}. We shall consider a two-parameter deformation of  (\ref{H}) where this symmetry is explicitly broken 
by coupling the electrons to a  chemical potential $\mu$ and  to a magnetic field $B$:
\bea\label{eq:twistedH}
H_{\mu,B}= H_0  - \mu { \hat{ N } } - 2 B S^{z}, 
\eea
with the conserved electron number ${ \hat{ N } }$ and spin $S^{z}$ operators defined as
\bea
{ \hat{N} } = \sum_{i=1}^L \left( c^\dagger_{i,\uparrow} c_{i,\uparrow} + c^\dagger_{i,\downarrow} c_{i,\downarrow} \right), \;\;\;\;\;
S^z = \frac{1}{2}  \sum_{i=1}^L \left(  c^\dagger_{i,\uparrow} c_{i,\uparrow} - c^\dagger_{i,\downarrow} c_{i,\downarrow} \right).
\eea
Lieb and Wu showed that this system is integrable and that it can be solved by means of the Bethe Ansatz (BA) method \cite{LiebWu}. The spectrum of the Hamiltonian
(\ref{eq:twistedH})  is characterized by two sets of complex quantum numbers (Bethe roots): the \emph{charge momenta} 
$\{ k_j\}_{j=1}^N $ and \emph{spin rapidities}, $\{ \lambda_j\}_{j=1}^M $. $N$ is the total number of electrons and $M$ is the number of 
down-spin electrons, with $2M \le N \le L$. 
The Bethe roots are solutions of the Lieb-Wu BA equations:
\begin{align} 
e^{ik_jL} &= \prod_{l=1}^{M} \left(\frac{\lambda_l-\sin(k_j) - i\bu}{\lambda_l-\sin(k_j) + i\bu}\right),\;\;  j \in \left\{1,\dots,N\right\}, \label{eq:Lieb-Wu}\\  
\prod_{j=1}^{N} \left(\frac{\lambda_l-\sin(k_j) - i\bu}{\lambda_l-\sin(k_j) + i\bu}\right) &=
\prod_{^{m=1}_{m\ne l} }^{M} \left(\frac{\lambda_l-\lambda_m - 2i\bu}{\lambda_l-\lambda_m + 2i\bu} \right),\;\;  l \in \left\{1,\dots,M\right\},\label{eq:Lieb-Wu1}
\end{align} 
and the corresponding energy is given by
\bea\label{eq:Betheenergy}
E = -2 \sum_{i=1}^N \cos(k_i) + \bu  \, ( L - 2 N ) - \mu \, N -  B \, ( N - 2 M ).
\eea

The system at finite temperature can be studied using the Thermodynamic Bethe Ansatz (TBA)  method~\cite{YangYang, Takahashi}. 
The TBA equations for the Hubbard model, an infinite set of coupled nonlinear integral equations, were first derived by Takahashi starting from   equations (\ref{eq:Lieb-Wu}),(\ref{eq:Lieb-Wu1}) and the classification of their solutions according to the so-called string hypothesis \cite{Takahashi}. 
An alternative approach was
introduced more recently by J\"uttner, Kl\"umper and Suzuki in \cite{DDVHubbard}, and is based on an associated integrable lattice model introduced by Shastry \cite{Shastry0, Shastry1} and the path integral approach to thermodynamics, also called Quantum Transfer Matrix (QTM) method
~\cite{Koma:1987, MSuzuki, SuzukiInoue, Suzuki:1990}. 
One of the main results of \cite{DDVHubbard} 
is a  much simpler set 
of only three coupled nonlinear integral equations (NLIEs) of Kl\"umper-Batchelor-Pearce-Destri-DeVega type~\cite{KBP,KlumperPearce0, KlumperPearce, DDV}. 
Although the two approaches lead to equivalent expressions for the Gibbs free energy, a  direct link between the two sets of nonlinear integral equations was until now not found for the Hubbard model. For the discussion of the equivalence between QTM and TBA in other models, see \cite{Equivalence}. 

 In this paper we shall reconsider the Hubbard model TBA inspired by some notable recent results obtained studying the TBA equations for ${\mathcal N}{=}4$ Super Yang-Mills (SYM). 
 The TBA was introduced in the AdS/CFT context~\cite{BFT, GKKV, AF} to overcome the so-called wrapping problem~\cite{Wrapping} affecting the Beisert-Staudacher equations. Recently, this very complicated set of TBA equations was recast into the greatly simplified form of a nonlinear matrix Riemann-Hilbert problem: the Quantum Spectral Curve or $\bP \mu$-system~\cite{QSC, AdS5Long}. This new formulation, contrary to the TBA, treats the full spectrum on an equal footing. It has already led to an impressive number of perturbative and nonperturbative results~\cite{QSCAtWork,QCDPomeron,VolinMarboe}. 
A similar reduction was also recently obtained \cite{ABJMnoi2} in the case of the $\mathcal{N}{=}6$ Chern-Simons theory, and made possible the determination of the so-called slope and interpolating functions, both nontrivial nonperturbative quantities~\cite{GromovSlope}. The role of the Quantum Spectral Curve approach in the general integrable model framework is not  fully understood and one of the purposes of the present work is to investigate whether a similar structure arises also in the context of the Hubbard model.

One of the main achievements of this paper is the reduction of the TBA equations derived by Takahashi to a closed system  of functional relations, involving only four functions
$\bP_a^{ \ua}(z)$, $\bP_a^{ \ra }(z)$, ($a =\p, \m $), entire on a two-sheeted Riemann surface and characterised by a specific asymptotics.
The equations are
\bea
\bP_{\p}^{\ra}(z) { \widetilde \bP}^{\ra}_{\m}(z) - { \widetilde \bP}^{\ra}_{\p}(z) \bP^{\ra}_{\m}(z) &=&   -2 \;\sinh( 2  \sq(z) ), \label{eq:constraint3a}\\
\bP^{\ua}_{\p}(z) { \widetilde \bP}^{\ua}_{\m}(z) - { \widetilde \bP}^{\ua}_{\p}(z) \bP^{\ua}_{\m}(z) &=& \,2  \; \sinh(  2  \sq(z) ), \label{eq:constraint3b}\\
\wT_{1,1}^{\ra}(z)  &=& e^{-\sq( z + i \bu ) + \sq(z - i \bu ) } \; \wT_{1,1}^{\ua}(z),\label{eq:constraint3c}
\eea
where $T \sq(z) = -i z \sqrt{ 1 - 1/z^2 } $, and 
\bea
\wT^{\alpha}_{1,1}(z) = \bP^{\alpha}_{\p}(z + i \bu) \, \bP^{\alpha}_{\m}(z - i \bu) - \bP^{\alpha}_{\m}(z + i \bu ) \, \bP^{\alpha}_{\p}( z - i \bu ), \;\;\; \alpha=\bigra, \bigua.
\eea
In (\ref{eq:constraint3a}) and (\ref{eq:constraint3b}), ${ \widetilde \bP }_i^{\alpha}$ denotes the second-sheet evaluation of $\bP_i^{\alpha}$. 
The Gibbs free energy $f$ is contained in the large-$z$ asymptotics, for instance
\bea\label{eq:finT}
\ln \wT^{\ra}_{1,1}(z)  &\sim &  -\frac{1}{T} ( f + \mu + \bu ).  
\eea
We also found an alternative formulation, based on the fact that the $\bP$ functions satisfy the functional relation
\bea\label{eq:new0}
{ \widetilde \bP }_a^{\ra}(z) \bP^{\ua}_b(z) &=& \bF_{ab}(z + i \bu) e^{ \sq(z) } +  \bF_{ab}(z - i \bu) e^{- \sq(z) },
\eea
where $\bF_{ab}(z)$ ($a, b = \p, \m$) are entire functions analytic on the whole complex plane. 
 The set of equations (\ref{eq:new0}) and (\ref{eq:constraint3a}),(\ref{eq:constraint3b}) is equivalent to (\ref{eq:constraint3a})-(\ref{eq:constraint3c}), and may be seen as the analogue of the AdS/CFT $\bP\mu$-system. 
 
 Furthermore, these relations imply that the zeros of the $\bP$ and $\bF$ functions are constrained by a set of exact Bethe Ansatz equations. More precisely, setting 
 ${\mathcal{ Q }_{\p} }(z) = e^{ i \hB z/\bu } \, \bP_{\p}^{\ra}(z) \, { \widetilde \bP }_{\p}^{\ra}(z) $ and $\f_{\p\m}(z) =  e^{ \frac{i}{2} (\hB - \hmu ) z / \bu  } \,\bF_{\p\m}(z)$, 
 we have
\bea
e^{ \hB - \hmu }  \; \frac{ \f_{\p\m}( s_i + i \bu )}{ \f_{\p\m}(s_i - i \bu ) } &=& - e^{ \beps( s_i )/T }, \;\;\; \text{at}\;\; \mathcal{ Q}_{\p}(s_i) = 0, \label{eq:BA00}\\
e^{-2 \hmu  } \;  \frac{ \f_{\p\m}( w_{\alpha} + 2 i \bu ) }{ \f_{\p\m}( w_{\alpha} - 2 i \bu ) } &=& -\frac{ \mathcal{ Q}_{\p}( w_{\alpha} + i \bu) }{\mathcal{ Q}_{\p}( w_{\alpha} - i \bu) }, \;\;\; \text{at} \;\; \f_{\p\m}( w_{\alpha} ) = 0.
\label{eq:BA0}
\eea
where $\hmu = \mu/T$, $\hB=B/T$ and $ \beps(z) = -2 T \sq(z) = 2 i z \sqrt{ 1 - 1/z^2 }$. 
Notice that the solutions of these equations have an infinite number of Bethe roots.
 In fact, (\ref{eq:BA00}), (\ref{eq:BA0}) coincide with the infinite Trotter number limit  of the exact Bethe Ansatz diagonalising the Quantum Transfer Matrix. Since the latter equations are the  starting point for the derivation of the NLIEs of \cite{DDVHubbard}, our analysis provides the missing link between the 
two different approaches to the Hubbard model thermodynamics. As mentioned above, this was a  longstanding problem (see, for example, the discussion at the end of Chapter 13 of \cite{HubbardBook}).

 In this paper we also present a preliminary study of the numerical solution method for the ground state. At least for sufficiently high values of the temperature, we expect that the ground state is singled out by the requirement that all the roots $s_i$ are located on the second sheet. In the region of validity of this assumption, our functional relations give rise to a simple set of nonlinear integral equations, easy to solve numerically. We have explicitly checked the agreement of our results against the TBA predictions for many points in the region $|B |<1$, $ |\mu| <1$, $0 < \bu < 2$ and $T \geq 1 $.  However, the study of \cite{DDVHubbard} indicates that there is a flow of roots from the second to the first sheet as $T$ is decreased. It should be possible to generalise our numerical method to this parameter region, as well as to excited states, but we leave this problem for future studies.
 
 As a last remark, we point out that this infinite Bethe Ansatz is reminiscent of the one 
 discovered in the AdS/CFT context 
 in \cite{GromovSever, Cusp2} (however, in the latter case a relation with a lattice construction is not known).

The rest of the paper is organised as follows. Section \ref{sec:TBA} contains the TBA equations written in a slightly modified form that highlights some of the symmetries of the model and can be implemented numerically without range restrictions on the parameters $\mu$  and $B$. 
Section \ref{sec:Ysystem} contains the  Y-system~\cite{ZamoYsystem} and the basic discontinuity  relations~\cite{Extended}  that allow its  extension   
to arbitrary branches  of the Riemann surface on which the Y functions, solution of the TBA equations, are defined. 
In Section \ref{sec:Tsystem},  taking hints from the recent results 
in AdS/CFT and motivated by the symmetry of the TBA equations, the Y's 
are parametrised in terms of T functions in two  alternative  gauges ($\wT^{\ua}$ and $\wT^{\ra}$). 
The simple analytic properties in these two gauges motivate  a further reparametrisation of the T's as $2\times 2$  determinants of more 
elementary objects: the $\bP$ functions. Moreover, a resolvent-type representation for the $\bP$'s 
allows to express all the relevant  quantities
in terms of  a  pair of densities $\rho_{\ua}$ and $\rho_{\ra}$ fulfilling a new set of NLIEs  defined on a finite support (described in Section \ref{sec:numerics}). 
Equation (\ref{eq:new0}) and the exact Bethe Ansatz equations are derived in Section \ref{sec:QTM}, where the connection with the results of \cite{DDVHubbard} is also discussed. The free-fermion limit is  briefly  discussed in Section \ref{sec:freefermion}. 
Section \ref{sec:numerics} contains the new NLIEs together with preliminary  numerics and a  brief description of the  numerical  technique  adopted.
Section \ref{sec:mirror} describes the formal transformation relating the thermodynamic equations to the finite-size ones describing the Hubbard chain with twisted boundary conditions. 
The  more technical parts of the  analysis are confined to four Appendices. 
 A proof of the special analytic properties of the Y functions on the Riemann section  with only short cuts (the ``magic'' sheet of \cite{FiNLIE}) is reported in Appendix \ref{sec:magic}. 
The study of the monodromy properties  of the $\bP$ functions and the proof that they 
live on a two-sheet Riemann surface is given in Appendix \ref{app:twosheets}. 
The derivation of the factors connecting the $\wT^{\ua}$ to the   $\wT^{\ra}$ gauge, 
the proof of equation (\ref{eq:constraint3c}) and the derivation of the  novel set of NLIEs are 
the main results of  Appendix \ref{app:proof}. 
Finally, Appendix \ref{app:dictionary} contains  a dictionary linking this work to the paper \cite{DDVHubbard}.  
\section{Thermodynamic Bethe Ansatz equations}
\label{sec:TBA}
\begin{figure}[t]
\begin{minipage}[b]{0.49\linewidth}
\raggedleft
\includegraphics[width=6.6cm]{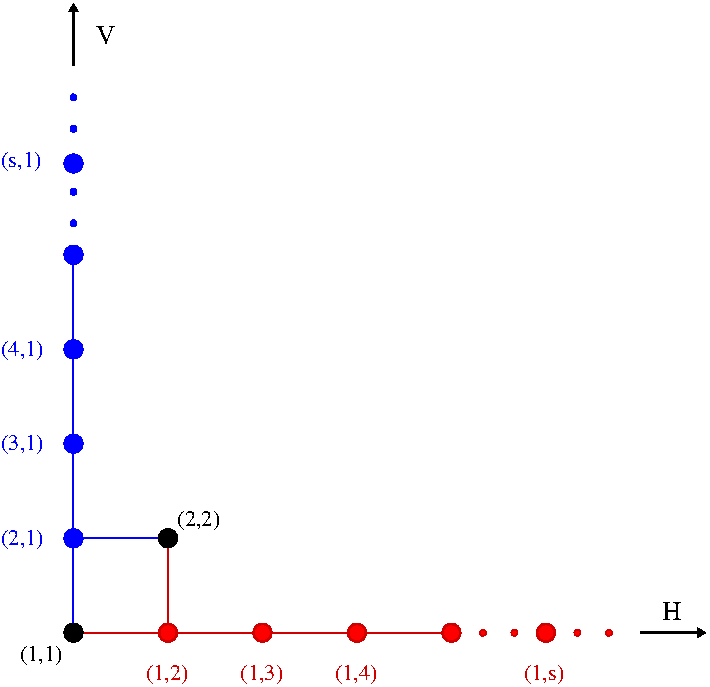}
\caption{ \label{fig:Ysystem} \footnotesize The diagram on which the Y-system (\ref{eq:Ysys}) is defined. }
\end{minipage}
\hspace{0.2cm}
\begin{minipage}[b]{0.49\linewidth}
\raggedleft
\includegraphics[width=7.cm]{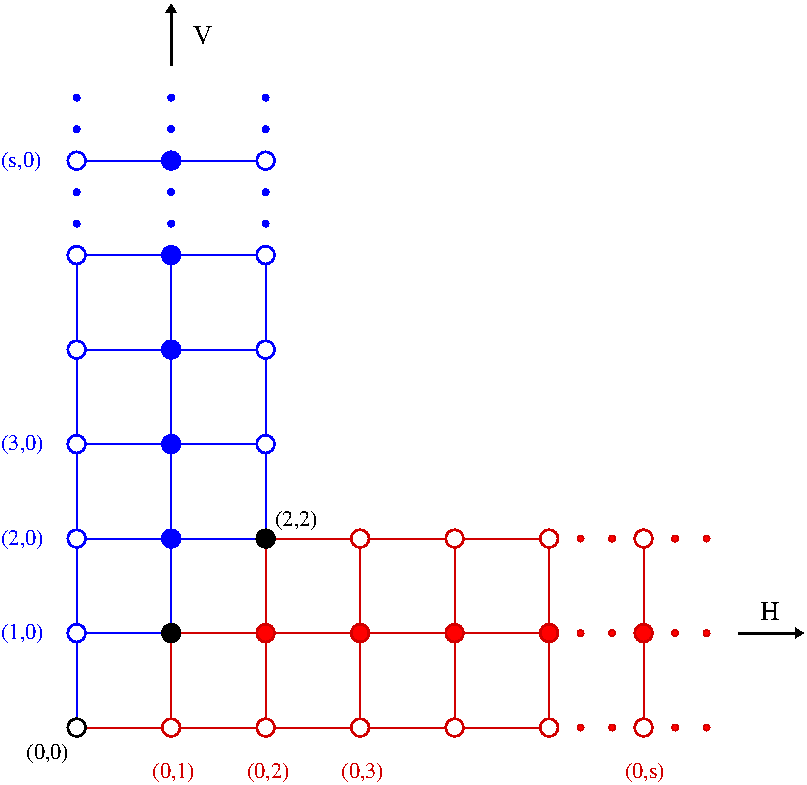}
\caption{ \label{fig:Tsystem} \footnotesize The extended diagram carrying the indices of the T functions. }
\end{minipage}
\end{figure}

In the notation of Takahashi, the solutions of the TBA equations are 
\bea
\left\{  \eta_n(z), \eta'_n(z), \zeta(k) \right\},  \;\; n \in \mathbb{N}^+,
\eea
where the physical range for the arguments is $z \in \mathbb{R}$ and $ k \in \left[-\pi, \pi\right]$. It is very convenient to 
reparametrise the variable $k$ as $z = \sin(k)$, and introduce the two-indexed functions $Y_{m, n}$ as follows: 
\bea
Y_{1,n}(z) &=& \eta_{n-1}(z), \;\;\;\; Y_{n,1}(z) = 1/\eta'_{n-1}(z), \;\;n \in \mathbb{N}^+,\\
Y_{1,1}( z ) &=& 1/\zeta( k ), \;\;\text{ for } k \in \left[-\frac{\pi}{2}, \frac{\pi}{2} \right],\\
Y_{2,2}(z) &=& \zeta(k), \;\; \text{ for }k \in \left[-\pi, -\frac{\pi}{2}\right] \cup\left[ \frac{\pi}{2}, \pi \right].
\eea
The structure of the TBA equations is thus codified on the L-shaped diagram represented in Figure \ref{fig:Ysystem}, with every node of the diagram associated to one of the unknown Y functions of the TBA. 
As a consequence of the change of variable $z = \sin(k)$, the functions $Y_{1,1}(z)$ and $1/Y_{2,2}(z)$ have a branch cut of square root type on the real axis for $z \in (-1, 1)$, and are in fact two branches of the same function $\zeta(z)$:  
 denoting with a tilde the analytic continuation around one of the branch points $z = \pm 1$, we have $
{ \widetilde Y }_{1,1}(z) = 1/Y_{2,2}(z)$. 

The TBA equations consistent with the Hamiltonian defined in (\ref{eq:twistedH}) can be written as
\bea
\log Y_{n+1, 1}(z) &=& 
 \int_{-1}^1\log \left(\frac{1 + Y_{2,2}(v) }{ 1+ 1/Y_{1,1}(v) } \right)\,a_n(v-z) \; dv - \sum_{m=1}^{\infty} \log \left(\frac{ 1 + Y_{m+1, 1} }{1+ 1/Z_{m+1}(\hmu) } 
\right)\ast a_{m, n} (z)  \nn\\
&-&\log{  Z_{n+1}(\hmu)}, \label{eq:tba1}
\eea
\bea
\log Y_{1, n+1}(z) &=&- \int_{-1}^1 \log \left(\frac{1 + 1/Y_{2,2}(v) }{ 1+ Y_{1,1}(v) } \right) \,a_n(v-z) \; dv + \sum_{m=1}^{\infty} 
\log \left(\frac{1 + 1/Y_{1, m+1} }{1 + 1/Z_{m+1}(\hB) } \right) \ast a_{m, n} (z)  \nn\\
&+&\log{ Z_{n+1}(\hB) } \label{eq:tba2},
\eea
\bea
\log Y_{2,2}(z) &=& -\log Y_{1,1}(z) + 2 \; \beps(z)/T = \sum_{m=1}^{\infty}  \left(\log \left(\frac{1 + Y_{m+1,1} }{1 + 1/Z_{m+1}(\hmu) } \right) -\log \left(\frac{ 1+ 1/Y_{1,m+1}}{1 + 1/Z_{m+1}(\hB)  } \right) \right) \ast a_n(z) \nn\\
&+&\log Z(\hB, \hmu) + \beps(z)/T - 2\; \bu/T, \label{eq:tba3}
\eea
where we have defined 
\bea
\hB &=& B/T, \;\;\;\; \hmu = \mu/T, \;\;\;\; \beps(z) = 2 i z \sqrt{ 1 - 1/z^2 } = -2 T \phi(z),\\
a_n(z) &=& \frac{n\bu}{\pi(n^2\bu^2 + z^2)},\;\;\;\;\;\; a_{m, n}(z) = \sum_{j=1}^n \left( a_{m+n-2 j}(z) + a_{m-n+2 j}(z) \right),  
\\
Z(\hB, \hmu) &=& \frac{\cosh(\hmu)}{\cosh(\hB) },\;\;\;  Z_n(x) = \frac{\sinh^2( n x )}{\sinh^2(x)} - 1.
\eea
In (\ref{eq:tba1})-(\ref{eq:tba3}), the convention for the convolutions is 
$a\ast b(z) = \InR dv \, a(v) b( v - z )$.
 Furthermore, here and in the rest of the paper we are implicitly assuming that the integrals over the interval $z \in (-1,1)$ run slightly above the branch cut. \\
 Notice that the TBA equations (\ref{eq:tba1})-(\ref{eq:tba3}) are written in a slightly different form as compared to \cite{Takahashi}. 
 For the equations of \cite{Takahashi}, the standard method of iterative solution~\cite{KlassenMelzer} requires sign restrictions on $B$ and $\mu$, while here this can be avoided since the symmetries  $B \lra -B $ and $\mu \lra -\mu$ of the ground state solution are explicit. 
 A further symmetry corresponds to the exchange of the two wings 
\bea\label{eq:verthor}
Y_{a, b} \lra 1/Y_{b, a},  \;\;\;\; B \lra \mu, 
\eea
 up to a change of sign for all the driving terms ($T \lra -T, \phi \lra -\phi$). \\
There are many equivalent expressions for the Gibbs free energy, for example: 
\bea\label{eq:energyrw}
f/T + \hmu &=& \bu/T - \log(2 \cosh(\hmu))  -\frac{1}{2\pi }\int_{-1}^{1} \log\left(( 1 + Y_{1, 1}(z) )( 1 + 1/Y_{2,2}(z))\right) \thh'(z) \, dz \nn\\
&&- \frac{1}{2\pi} \sum_{ n = 1 }^{\infty}\InR \log \left(\frac{ 1 + Y_{n+1,1}(z) }{ 1 + 1/Z_{n+1}(\hmu) } \right) \, \thh'_n(z) \; dz,
\eea
where
\bea 
\thh(z) &=& \arcsin( z ) =  i \log( -\x( z ) ), \;\;\;\; \x(z) = iz + iz \sqrt{1-1/z^2},\;\;\;\;
\thh'(z) = \frac{d}{dz} \thh(z) = \frac{i}{ z \sqrt{1-1/z^2} },  \label{eq:Zhukovsky} \nn\\
\thh_n(z) &=& \thh(z+i n \bu) - \thh(z-i n \bu),
\eea
so that
$\thh'_n(z) =  \frac{1}{\sqrt{ 1 - ( z + i n \bu  )^2 } } + \frac{1}{\sqrt{ 1 - ( z- i n \bu  )^2 } }$. 
The free energy can equivalently be rewritten using only the Y functions of the horizontal part of the diagram of Figure \ref{fig:Ysystem}, as
\bea\label{eq:energyuw}
f/T + \hmu  &=&- \bu/T -\log(2 \cosh(\hB)) -\frac{1}{2\pi }\int_{-1}^{1} \log\left(( 1 + 1/{Y_{1, 1}(z)} )( 1 + Y_{2,2}(z) )\right) \thh'(z) \, dz \nn\\
&&- \frac{1}{2\pi} \sum_{ n =1 }^{\infty} \InR \log \left(\frac{ 1 + 1/Y_{1,n+1}(z) }{ 1 + 1/Z_{n+1}(\hB) } \right) \, \thh'_n(z) \; dz.
\eea
The equivalence between (\ref{eq:energyrw}) and (\ref{eq:energyuw}) can be checked by using the TBA equation (\ref{eq:tba3}), and reflects the symmetry (\ref{eq:verthor}) and the properties~\cite{HubbardBook}:
\bea
f( -B, \mu, T) = f(B, \mu, T), \;\;\;\; f(B, -\mu, T) = f(B, \mu, T) + 2 \mu.  
\eea
Equations describing excited branches of the free energy -- which control the correlation lengths among local operators at finite temperature -- can in principle be obtained by analytic continuation~\cite{DT}, see also \cite{BLZExcitedStates, DoreyPerturbedCFT, FendleyIndices,FioravantiDDV}. 
For the Hubbard model, this has been accomplished only for a few states~\cite{HubbardTsunetsugu, klumper1996exact, HubbardCorrelation}. 
The TBA equations appear not to be the optimal tool for this analysis, and the results of \cite{HubbardTsunetsugu, klumper1996exact, HubbardCorrelation} were obtained adopting the QTM method.  
It would be important to have a more complete understanding of the free energy spectrum, 
and one of the aims of the present work is to propose an alternative formulation which seems to have some advantages over the existing approaches.
\begin{figure}[t]
\begin{minipage}[t]{0.49\linewidth}
\raggedleft
\includegraphics[width=6.6cm]{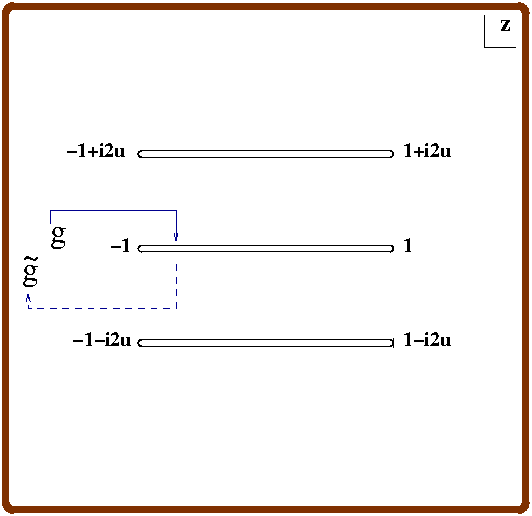}
\caption{ \footnotesize The ``magic'' sheet. }
\label{fig:magicsheet}
\end{minipage}
\hspace{0.2cm}
\begin{minipage}[t]{0.49\linewidth}
\raggedleft
\includegraphics[width=6.6cm]{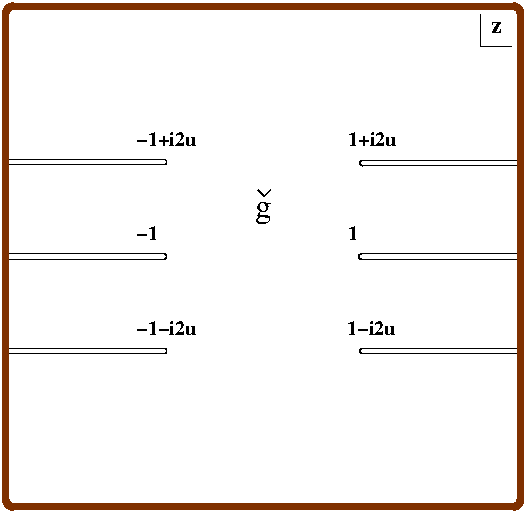}
\caption{ \footnotesize The ``mirror'' sheet, on which the Y-system and T-system are defined. }
\label{fig:mirrorsheet}
\end{minipage}
\end{figure}

\section{The Y-system and the discontinuity relations}
\label{sec:Ysystem}
 In this Section we shall describe the essential analytic properties of the system, necessary for the simplifications to be . 
 An important complication comes from the fact that the Y functions have square root branch points at certain
positions in the complex rapidity plane.  In particular, it can be seen analysing the TBA equations that $Y_{1, n}$ and $Y_{n, 1}$ are analytic in a strip $|\IIm(z)| < (n-1) \bu$, but have square root branch points at  
  $z = \pm 1 + i (n-1) \bu $, $z = \pm 1 - i (n-1) \bu $, and possibly at other positions further from the real axis. \\
To describe this structure, let us setup some useful notation. We adopt the convention that, for complex $z$, the Y's are evaluated on a Riemann sheet with short cuts, the ``magic'' sheet (see Figure \ref{fig:magicsheet}), and use the notation $\widetilde{g}(z)$ for the analytic continuation of a function $g(z)$ around either\footnote{It can be checked that the result of the analytic continuation around two branch points symmetric with respect to the imaginary axis is always the same.} of the branch points $z = \pm 1$. Since we reserve this notation to branch points of square root type, we always have the property ${\widetilde { \widetilde g }}=g$. Another notation that we shall use often is 
$g^{[+n]}(z) = g(z + i n \bu)$, where the shifts are evaluated on the magic sheet. 

\begin{figure}
\begin{minipage}[b]{0.5\linewidth}
\raggedleft
\includegraphics[width=7.8cm]{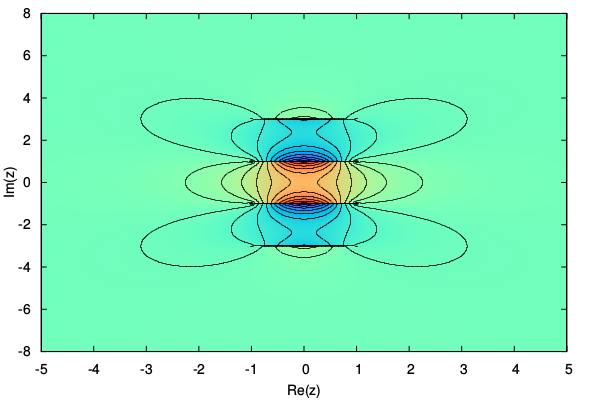}
\caption{ \label{fig:C3} \footnotesize A contour plot for $\TPsi^{(-)}_{2,1}(z)$ in the complex z-plane
confirms the presence of only four short cuts for $Y_{2,1}(z)$ on the magic sheet.
($\bu=1$, $T=0.5$, $B=0.4$, $\mu=0.1$)}
\end{minipage}
\hspace{0.0cm}
\begin{minipage}[b]{0.5\linewidth}
\raggedleft
\includegraphics[width=7.8cm]{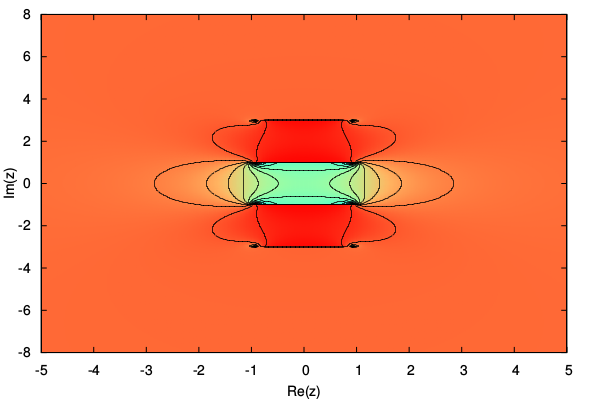}
\caption{ \label{fig:C4} \footnotesize  A contour plot for $\TPsi^{(+)}_{1,2}(z)$ in the complex z-plane 
confirms  the presence of only four short cuts for $Y_{1,2}(z)$ on the magic sheet.
($\bu=1$, $T=0.5$, $B=0.4$, $\mu=0.1$)}
\end{minipage}
\end{figure}

From (\ref{eq:tba1})-(\ref{eq:tba3}) with $n=1$, one can derive the following discontinuity relations~\cite{Extended,ABJMdisco}  for the monodromies around some of the branch points closest to the real axis
\bea
{ \widetilde Y }_{1,1}(z) &=& 1/Y_{2,2}(z), \;\;\;\; 
{Y_{1,1}(z) Y_{2,2}(z) } = e^{-4 \sq(z)}, \label{eq:disco1}\\
 Y_{2,1}^{[+1]}(z)/{ \widetilde{ Y^{[+1]}_{2,1}  }}(z) &=& \left(\frac{1 + 1/Y_{1,1}(z) }{1 + Y_{2,2}(z)}\right), \label{eq:disnew} \\
 Y_{1,2}^{[+1]}(z)/{ \widetilde{ Y^{[+1]}_{1,2}  }}(z)  &=& \left(\frac{1 +  1/Y_{2,2}(z)}{1 + Y_{1,1}(z)}\right). \label{eq:dis4}
\eea
 Furthermore, a crucial property implied by the TBA equations is that the Y's are solutions of the  
Y-system~\cite{ZamoYsystem}
\bea\label{eq:Ysys}
{ \check{Y} }_{m,n}(z+i \bu ) \; { \check{Y} }_{m,n}(z - i \bu) = \frac{( 1 + { \check{Y} }_{m,n+1}(z) )( 1 + { \check{Y} }_{m,n-1}(z) )}{ ( 1 + 1/{ \check{Y} }_{m+1,n}(z) )( 1 + 1/{ \check{Y} }_{m-1,n}(z) ) } ,
\eea
where $(m, n) \in \left\{ (1, k), k\in \mathbb{N}^+ \right\} \cup \left\{ (k, 1), k\in \mathbb{N}^+\right\} $ with boundary conditions 
$Y_{k, 0}=1/Y_{0, k} = Y_{k+2, 2}=1/Y_{2, k+2} = 0$ for $k \in \mathbb{N}^+$ (see Figure \ref{fig:Ysystem}). Notice that there is no independent equation centered at the node $(2,2)$. 
Indeed, the function $Y_{2,2}$ is simply related to $Y_{1,1}$ through equations (\ref{eq:disco1}). 
In (\ref{eq:Ysys}), we have used the notation ${\check{Y} }$ to emphasise that the functional relations (\ref{eq:Ysys}) are valid 
 on the specific Riemann section shown in Figure \ref{fig:mirrorsheet}, the ``mirror'' sheet, where all the branch cuts are traced as semi-infinite lines parallel to the real axis, and do not cross the strip $|\RRe(z)| < 1$. This gives a precise prescription on how to evaluate the shifted values appearing on the lhs of (\ref{eq:Ysys}). 
 Adapting the arguments of ~\cite{Extended,ABJMdisco,BalogHegedus}, it can be proved that the functional relations (\ref{eq:disco1})-(\ref{eq:Ysys}), together with the asymptotics 
\bea\label{eq:Yasy}
\log\left(\frac{1 + 1/Y_{1,n+1}(z) }{1 + 1/Z_{n+1}(\hB) }\right) &=& \text{O}\left(\frac{1}{z^2} \right),\;\;\; \log\left(\frac{1 + Y_{n+1,1}(z) }{1 + 1/Z_{n+1}(\hmu) }\right) = \text{O}\left(\frac{1}{z^2} \right), \;\;\; n \in \mathbb{N}^+, \\
\log Y_{2,2}(z) &=&  \log Z(\hB,\hmu) + \beps(z)/T -2 \bu/T + \text{O}\left(\frac{1}{z^2} \right),
\eea
and the assumption that the Y's have no zeros or poles in the strip $|\IIm(z)| \leq \bu$, are fully equivalent to the TBA. \\
One might expect that 
each Y function should display, on a generic Riemann section, an infinite ladder of further square root branch points at steps of $2 i \bu$, replicated from the ones closer to the real axis by the Y-system (\ref{eq:Ysys})~\cite{Extended}. 
However, this is not the case. We shall indeed prove that each of the Y functions has only a finite number of branch points on any sheet and that the number of Riemann sheets is actually finite. 
The fact that all functions appearing in this problem are defined on a finite genus Riemann surface is perhaps obvious from the perspective of the exact Bethe Ansatz and the Quantum Transfer Matrix construction of \cite{DDVHubbard}. However, starting from the TBA equations this property is much harder to prove. To establish this result, we adopt the following strategy. First, we show that the Y functions have at most four branch cuts on the magic sheet:
\begin{itemize}
\item  $Y_{1,n}(z)$ and $Y_{n,1}(z)$  for $n \geq 2$ have only four branch cuts at
$$ 
z \in(-1,1) \pm i \bu (n-1), \;\;\;\; z \in (-1,1) \pm i \bu (n+1),
$$ 
\item $Y_{1,1}(z) $ and $\hY_{2,2}(z)$ have only three branch cuts at
$$ 
z \in (-1,1), \;\;\;\; z \in (-1,1) \pm 2 i \bu.
$$
\end{itemize}
This result is established in Appendix~\ref{sec:magic}\footnote{It is important to notice that  the proposed proof can be  straightforwardly adapted to the known  AdS/CFT cases, this gives  an alternative and perhaps   more transparent way to understand the nice  properties  discovered in \cite{FiNLIE} for the AdS$_5$/CFT$_4$ Y functions on the magic sheet.}.  
Secondly, 
it is shown in Appendix \ref{app:twosheets} that no further branch cuts can appear on the other sheets. 
A direct numerical solution of the TBA equations  highlights, unequivocally,  the presence of at most four short cuts for the functions $Y_{1,n}$ and $Y_{n,1}$, both on the magic and mirror sections.
For the magic sheet, contour plots for the functions  $\TPsi^{(+)}_{12}$ and $\TPsi^{(-)}_{21}$ with 
\bea
\TPsi^{(\pm)}_{a,b}(z)= \frac{\left| 1 +  (Y_{a,b}(z))^{\pm 1} \right|}{1+\left| 1 +  (Y_{a,b}(z))^{\pm 1} \right|}
\eea
are displayed in  Figures \ref{fig:C3} and \ref{fig:C4}.  

\section{The T-system}
\label{sec:Tsystem}
Following the strategy which allowed the simplification of the AdS/CFT TBA, 
we  shall now make a chain of simplifications and reduce relations (\ref{eq:disco1})-(\ref{eq:Ysys}) to the finite set of constraints (\ref{eq:constraint3a})-(\ref{eq:constraint3c}). 
The first step is to parametrise the Y functions as
\bea\label{eq:YtoT}
Y_{n,s}(z) = \frac{ T_{n,s+1}(z) T_{n,s-1}(z) }{ T_{n+1,s}(z) T_{n-1,s}(z) }, \;\;\;\;n,s \in \mathbb{N}^+,
\eea
 where the T's ($T_{n,s}$ with $n,s \in \mathbb{N}$) are defined on the extended lattice displayed in Figure \ref{fig:Tsystem} and satisfy, on the mirror section, a discrete Hirota equation (the T-system) 
\bea\label{eq:Tsystem}
\check{T}_{n,s}( z + i \bu ) \check{T}_{n,s}( z - i \bu ) = \check{T}_{n + 1,s }(z) \check{T}_{n-1,s}(z) +  \check{T}_{n,s-1}(z)  \check{T}_{n,s+1}(z),
\eea
with boundary conditions 
$\check{T}_{k-1, -1}= \check{T}_{-1, k-1} = \check{T}_{k+2, 3}=\check{T}_{3, k+2} = 0$ for $k \in \mathbb{N}^+$. 
Notice that the Y-system  (\ref{eq:Ysys}) is automatically fulfilled as a consequence of (\ref{eq:YtoT}) and (\ref{eq:Tsystem}). 
However, the parametrisation (\ref{eq:YtoT}) is not one-to-one. Different solutions of the T-system, corresponding to the same set of Y's, are linked by a gauge transformation~\cite{WiegmannZabrodin}. 
 A clever gauge choice can greatly simplify the problem. 
\subsection{Vertical and horizontal gauges}
We shall now introduce the two gauges $\wT^{\ra}$ and $\wT^{\ua}$, each defined by a set of simple conditions on one of the wings, horizontal ($\textsf{H}$) or vertical ($\textsf{V}$), of the diagram in Figure \ref{fig:Tsystem}. 
 We require that they fulfill the following properties, compatible with the cut structure of the Y functions, summarised in Section  \ref{sec:Ysystem}: %
  \bea
&&  \wT^{\ra}_{1,n}(z), (n \geq 1) \;\;\; \text{ has only two branch cuts at } z \in (-1,1) \pm i n \bu \label{eq:cutsTr}\\
 && \wT^{\ra}_{0,n}(z) = 1, \;\;\;\;\;\; \wT^{\ra}_{2,n}(z) = \wT^{\ra\,[+n]}_{1, 1}( z ) \, \wT^{\ra\,[-n]}_{1, 1}(z), \;\;(n \geq 2)\label{eq:cutsTr2}
 \eea
and
\bea
&&  \wT^{\ua}_{n,1}(z), (n \geq 1) \;\;\; \text{ has only two branch cuts at } z \in (-1, 1) \pm i n \bu \label{eq:cutsTu} \\
 && \wT^{\ua}_{n, 0}(z) = 1, \;\;\;\;\;\; \wT^{\ua}_{n,2}(z) = \wT^{\ua \, [+n]}_{1, 1}( z) \, \wT^{\ua \,[-n]}_{1,1}(z), \;\;(n \geq 2).\label{eq:cutsTu2}
 \eea
The conditions listed above still leave a residual gauge invariance. These gauges are fixed uniquely by imposing the following extra conditions:
\bea\label{eq:T10wr}
\wT_{1,0}^{\ra}(z) = e^{-2 \phi(z)}, \;\;\;\;\; \wT_{0, 1}^{\ua}(z) =  e^{ 2  \phi(z) }.
\eea
It is shown in Appendix \ref{app:twosheets} that this choice is indeed possible. 
Given these definitions, it is far from obvious that also the functions $\wT^{\ra}_{n, m}$ with $n > m$    and $\wT^{\ua}_{n, m}$ with $m > n$ have such a simple cut structure. 
However, one of our main results, is that these two gauges are in fact connected by an elementary transformation:
\bea\label{eq:wTgen}
\frac{\wT_{1,s}^{\ra}(z)}{\wT_{1,s}^{\ua}(z)} &=& \frac{\wT_{s, 1}^{\ra}(z) }{\wT_{s, 1}^{\ua}(z)}= e^{-  \phi^{[+s]}(z) + \phi^{[-s]}(z) }, \, \;\; s \in \mathbb{N}^+, \label{eq:wTgen1}\\
\wT_{s,0}^{\ra}(z) &=& \frac{1}{\wT_{0,s}^{\ua}(z)} = e^{-2 \sum_{n = 0}^{s-1}  \text{sgn}(-s+1 + 2 n) \; \phi^{[-s+1 + 2 n ]}(z)  }, \,\,\,\,\, s \in \mathbb{N}, \label{eq:wTgen2}\\
\frac{\wT_{2,s}^{\ra}(z)}{\wT_{2,s}^{\ua}(z)} &=& \frac{\wT_{s, 2}^{\ra}(z) }{\wT_{s, 2}^{\ua}(z) }= e^{- \phi^{[s+1]}(z) + \phi^{[-s-1]}(z) + \phi^{[s-1]}(z) - \phi^{[1-s]}(z) }, \;\;\; s = 2, 3, \dots  \label{eq:wTgen3}
\eea
To prove (\ref{eq:wTgen1})-(\ref{eq:wTgen3}), it is necessary to parametrise (\ref{eq:cutsTr})-(\ref{eq:cutsTu2}) with more elementary building blocks, the $\bP$ functions introduced in Section \ref{sec:P} and the resolvent parametrisation of Section \ref{sec:resolvent}. The technical details of the derivation 
are described in Appendix \ref{app:proof}.

In the matching condition (\ref{eq:wTgen1})-(\ref{eq:wTgen3}), together with the requirement that the $\wT$'s have no poles, is hidden the full content of the original TBA equations, and its generalisation to excited states. 
Furthermore, we will see in Section \ref{sec:QTM} that the $\wT$ functions have a clear interpretation: they are directly related to the eigenvalues of the (fused) Quantum Transfer Matrix in the infinite Trotter number limit~\cite{DDVHubbard}.
\subsection{The $\bP$ functions}\label{sec:P}
The next natural step is to parametrise the $\wT$ functions defined in (\ref{eq:cutsTr})-(\ref{eq:T10wr}), and possessing only two cuts, in terms of  
 objects 
having only one branch cut running along $(-1, 1)$. We introduce a pair of $\bP_a$ ($a=\p,\m$) functions for each of the wings, and write 
\bea
\wT_{1,s}^{\ra}(z) &=& \bP_{\p}^{\ra \,[+s]}(z) \bP_{\m}^{\ra \,[-s]}(z) - \bP_{\p}^{\ra \,[-s]}(z)  \bP_{\m}^{\ra \,[+s]}(z), \label{eq:detr}\\
\wT_{s,1}^{\ua}(z) &=&\bP_{\p}^{\ua \,[+s]}(z) \bP_{\m}^{\ua \, [-s]}(z) -  \bP_{\p}^{\ua \,[-s]}(z) \bP_{\m}^{\ua \,[+s]}(z), \label{eq:detu}   
\eea
with $s \in \mathbb{N}^+$. 
Thanks to relations of Pl\"ucker's type among determinants, this  parametrisation  is automatically consistent with the T-system in the corresponding $\bigra$ or $\bigua$ wing of the diagram. 
This ``quantum Wronskian'' construction \cite{BLZ2,WiegmannZabrodin} suggests that the $\bP$'s are related to 
the Q functions describing the spectrum of the thermodynamic problem~\cite{DDVHubbard}; this relation is made precise in Section \ref{sec:QTM}.\\
To fix completely these functions, we have to specify their asymptotics and to constrain further their analytic properties. 
As suggested by the asymptotics of the Y functions (\ref{eq:Yasy}), we demand that\footnote{In analogy with the AdS/CFT case~\cite{AdS5Long}, for excited states we expect that this asymptotic behaviour may be modified by power-like prefactors encoding the quantum numbers.}, for large $z$,
\bea\label{eq:Pasy}
\bP^{\ra}_{\bpm}(z)  &\sim & \frac{  C_{\ra}  }{ \sqrt{ 2 \, \sinh ( \hB  ) } }\; e^{ \mp \frac{i}{2} \hB  z / \bu   }, 
\;\;\;\;\;
\bP^{\ua}_{\bpm}(z)  \sim   \frac{ C_{\ua} }{ \sqrt{ 2 \, \sinh( \hmu )  } } \; e^{  \mp \frac{i}{2} \hmu  z/\bu   }.
\eea
The particular normalization of the prefactors $C_{\ra}$ and $C_{\ua}$ in (\ref{eq:Pasy}) is chosen for future convenience. 
As shown in Appendix \ref{app:twosheets}, the four $\bP$ functions live on a two-sheet Riemann surface. This is an important simplification as compared to the AdS/CFT case studied in \cite{QSC}, where the analogous $\bP$ functions have additional infinitely many branch cuts, starting from their second sheet\footnote{ In AdS/CFT, the structure associated to the presence of a ladder of infinitely many branch cuts is encoded in a periodic matrix $\mu_{ij}$. In the current setup, the introduction of a periodic matrix can be avoided.}. 
 As we discuss in Appendix \ref{app:proof}, equation (\ref{eq:T10wr}) gives the constraint
\bea
\bP_{\p}^{\ra}(z) { \widetilde \bP}^{\ra}_{\m}(z) - { \widetilde \bP}^{\ra}_{\p}(z) \bP^{\ra}_{\m}(z) &=&   -2 \;\sinh( 2  \phi(z) ), \label{eq:constraint31}\\
\bP^{\ua}_{\p}(z) { \widetilde \bP}^{\ua}_{\m}(z) - { \widetilde \bP}^{\ua}_{\p}(z) \bP^{\ua}_{\m}(z) &=& 2  \; \sinh( 2  \phi(z)). \label{eq:constraint32}
\eea
Equations (\ref{eq:constraint31})-(\ref{eq:constraint32}), together with the relation giving the matching of the two wings:
\bea\label{eq:wT}
\wT_{1,1}^{\ra}(z) = e^{-\phi^{[+1]}(z) + \phi^{[-1]}(z) } \, \wT_{1,1}^{\ua}(z),
\eea
 the asymptotics (\ref{eq:Pasy}) and information on the number of zeros on the first Riemann sheet, form a closed set of conditions for the $\bP$ functions. In Section \ref{sec:numerics}, we will show how to solve these equations numerically. 
Notice that the constants $C_{\ra}$ and $C_{\ua}$ are fixed by the solution to this system, and contain the free energy $f$ through the relation:
\bea\label{eq:asyC}
f/T + \hmu = -2 \log C_{\ra}  -\bu/T  = -2 \log C_{\ua}  + \bu/T.
\eea
We will prove (\ref{eq:asyC}) in Section \ref{sec:freeenergy}, with the aid of a very convenient parametrisation in terms of resolvents, inspired by \cite{FiNLIE,Cusp2}. 
\subsection{Resolvent parametrisation}\label{sec:resolvent}
In the following, we shall assume that the $\bP$ functions have  no zeros on the first sheet. We expect that this singles out the ground state solution in a large parameter region for sufficiently high temperatures. 
 The numerical solution of the functional relations presented in Section \ref{sec:numerics} reveals the existence of an infinite number of zeros on the second sheet. Starting from the ground state, we expect that excited states can be obtained by analytic continuation in $B$ and $\mu$, 
 leading to the migration of a finite number of zeros to the first sheet and to simple modifications in the equations presented in this Section. 
  We point out that, in a neighbourhood of $T=0$, even the ground state solution should present a finite number of zeros on the first sheet~\cite{DDVHubbard}, and therefore the following analysis needs to be modified.  

Let us introduce two resolvent functions
\bea
\CG_{\alpha}(z) = \frac{1}{2 \pi i } \int_{-1}^{1} \frac{\brho_{\alpha}(v)}{z - v } dv, \;\;\;\alpha=\bigra, \bigua,
\eea
where the densities $\rho_{\ra}$ and $\rho_{\ua}$ are for the moment undetermined. This is the most generic parametrisation of a function with a single cut, no zeros and no poles on the first sheet, and large-$z$  asymptotics $\mG_{\alpha} \sim 1/z$ on the first sheet. Under analytic continuation through the cut, the resolvents transform as
 \bea\label{eq:integralG}
{ \widetilde  \CG }_{\alpha}(z) = \CG_{\alpha}(z) + \rho_{\alpha}(z).
 \eea
Thus, under the assumptions discussed above, we can parametrise the  $\bP$'s as
\bea\label{eq:resolventP}
 \sqrt{ 2 \, \sinh( \hB ) } \, \bP_{\bpm}^{\ra}(z) &=& \bh_{\ra}(z) \, e^{ \pm(\hB \mG_{\ra}(z)  - \frac{i}{2} \hB z / \bu )}, \\
 \sqrt{ 2 \, \sinh( \hmu ) } \, \bP_{\bpm}^{\ua}(z) &=& \bh_{\ua}(z) \, e^{ \pm (\hmu \mG_{\ua}(z)  - \frac{i}{2} \hmu z / \bu )}, \nn
\eea
where the functions $\bh_{\alpha}$ ($\alpha =\bigra, \bigua$) are required to have a single cut, no zeros or poles and  a constant leading asymptotics on the first sheet. 
Taking ratios of these functions we find 
\bea\label{eq:exprho0}
e^{2 \hB \brho_{\ra} } =  \frac{\tP_{\p}^{\ra} \bP_{\m}^{\ra}}{ \tP_{\m}^{\ra} \bP_{\p}^{\ra}},\;\;\;\; e^{2 \hmu \brho_{\ua} } = 
\frac{\tP_{\p}^{\ua} \bP_{\m}^{\ua}}{ \tP_{\m}^{\ua} \bP_{\p}^{\ua}}.
\eea
As a consequence of the simple monodromy properties of the $\bP$'s, also  $\rho_{\ra}$ and $\rho_{\ua}$ live on a two-sheet Riemann surface, with
\bea
{ \widetilde \rho }_{\alpha}(z) = - \rho_{\alpha}(z).
\eea
\begin{figure}[t]
\begin{minipage}[b]{0.5\linewidth}
\centering
\includegraphics[width=7.3cm]{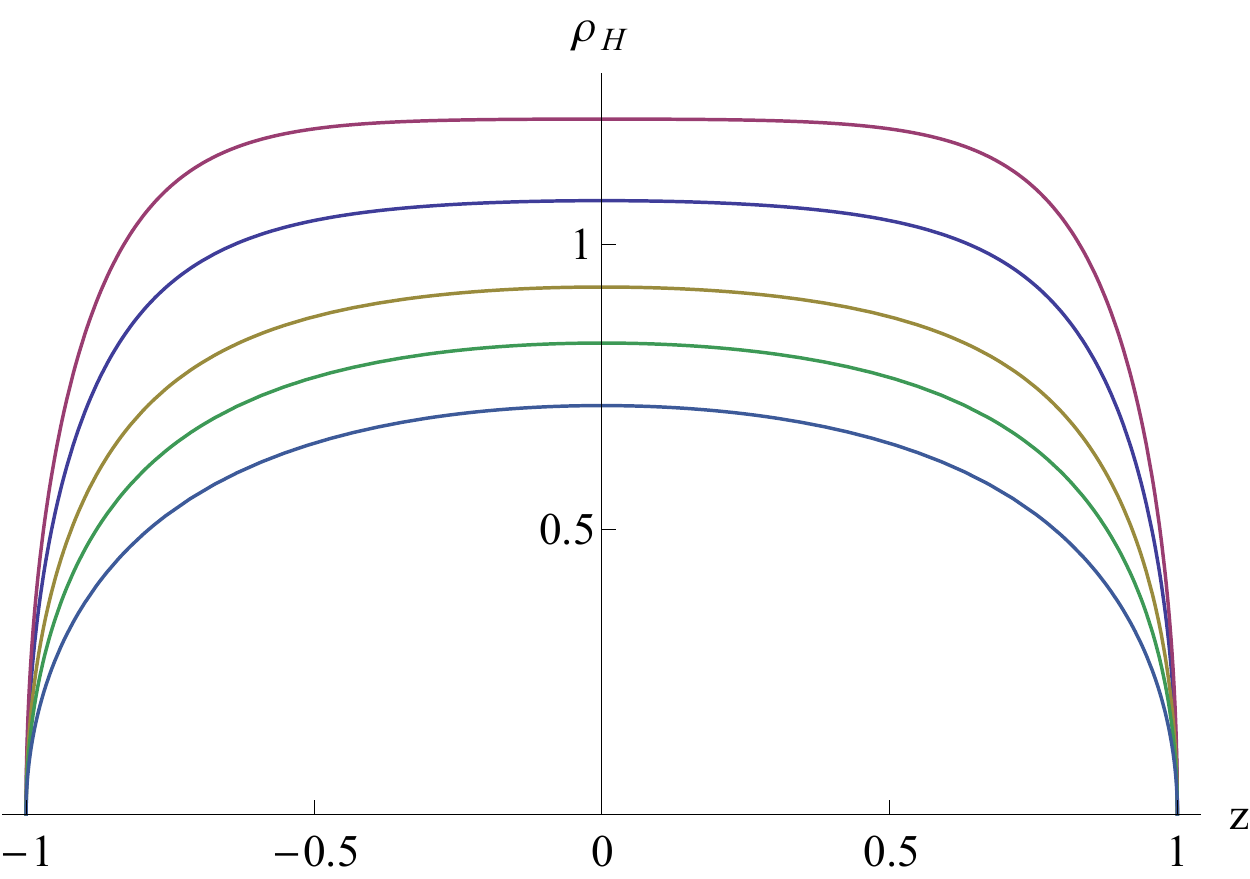}
\caption{ $\rho_{\ra}$ for $B=1$, $\mu=0.5$ and $T=1$, for $\bu = 0$, $\bu = 0.25$, $\bu=0.5$, $\bu = 1$ and $\bu = 2$.  
$\rho_{\ra}$ is monotonically increasing with $\bu$. }
\label{rho1}
\end{minipage}
\hspace{0.1cm}
\begin{minipage}[b]{0.5\linewidth}
\centering
\includegraphics[width=7.3cm]{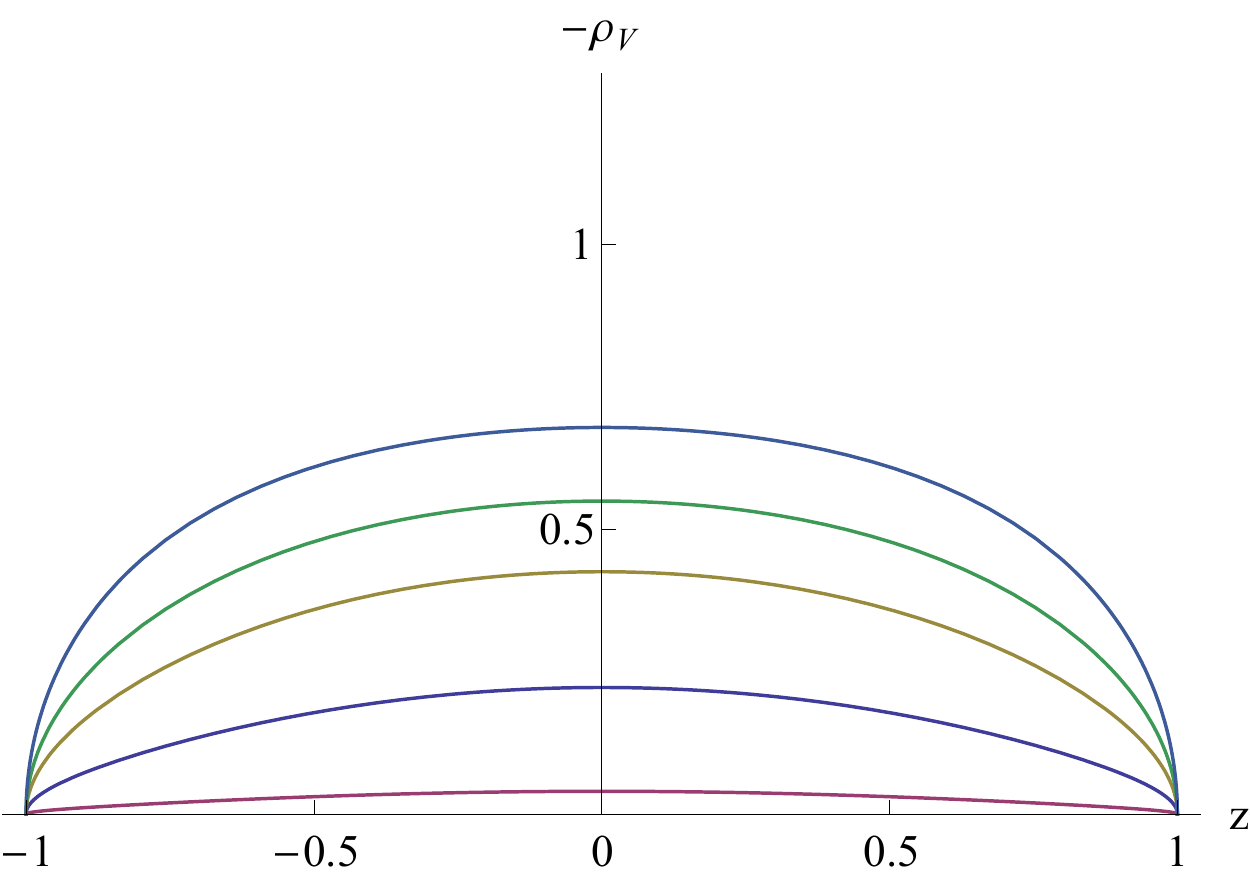}
\caption{$-\rho_{\ua}$ for $B=1$, $\mu=0.5$ and $T=1$, for $\bu = 0$, $\bu = 0.25$, $\bu=0.5$, $\bu = 1$ and $\bu = 2$. 
$-\rho_{\ua}$ is monotonically decreasing with $\bu$.}
\label{rho2}
\end{minipage}
\end{figure}
More precisely, we require that the behaviour at the branch points is such that the functions $\rho_{\alpha}(z)/(z \sqrt{ 1 - 1/z^2 })$ are analytic in a neighbourhood of the cut $z \in (-1, 1)$. Due to (\ref{eq:exprho0}), however, the densities can have logarithmic singularities elsewhere in the complex plane, in correspondence to zeros of the ${ \widetilde \bP}$'s.\\
Substituting (\ref{eq:resolventP}) into (\ref{eq:constraint31}) and (\ref{eq:constraint32}), we find 
\bea\label{eq:hhtnew2}
\bh_{\ra}(z) { \widetilde \bh }_{\ra}(z)  \sinh( \hB  \rho_{\ra}(z) ) &=& 2 \,\sinh( \hB) \, \sinh( 2 \phi(z)  ), \label{eq:hhtnew2a}\\
\bh_{\ua}(z) { \widetilde \bh }_{\ua}(z)  \sinh( \hmu \rho_{\ua}(z) ) &=& - 2 \, \sinh( \hmu) \, \sinh( 2 \phi(z) ).\label{eq:hhtnew2b}
\eea
 The solution of the latter equations, compatible with the asymptotics (\ref{eq:Pasy}) and the absence of zeros on the first sheet, is unique and given by\footnote{Notice that in (\ref{eq:integralh}), $\phi(v)$ needs to be evaluated just above the branch cut, so it agrees with $\frac{1}{T} \sqrt{1-v^2}$.}:
\bea\label{eq:integralh}
\log\bh_{\ra}(z) &=& -\frac{z \sqrt{ 1 - 1/z^2 }}{ 2 \pi } \; \int_{-1}^{1} \log\left(  \frac{ 2\sinh( 2 \phi(v) ) \sinh( \hB ) }{ \sinh(  \hB \rho_{\ra}(v) ) } \right)  \frac{ dv }{ \sqrt{1 - v^2 } \; ( v - z ) }, \\
\log\bh_{\ua}(z) &=& -\frac{z \sqrt{ 1 - 1/z^2 }}{ 2 \pi } \; \int_{-1}^{1} \log\left( - \frac{ 2 \sinh(  2 \phi(v) ) \sinh( \hmu )}{ \sinh(  \hmu \rho_{\ua}(v) ) } \right)  \frac{ dv }{ \sqrt{1 - v^2 } \; ( v - z ) }. \label{eq:integralh2}
\eea
Thus, with the definitions (\ref{eq:integralh}) and (\ref{eq:integralh2}), the parametrisation (\ref{eq:resolventP}) automatically fulfills the constraints (\ref{eq:constraint31}),(\ref{eq:constraint32}). \\
Considering the large-$z$ asymptotics of (\ref{eq:integralh}),(\ref{eq:integralh2}) we find that, on the first sheet,
\bea\label{eq:Cintegral}
\lim_{z \rightarrow \infty} \log\bh_{\ra}(z)
&=& \log C_{\ra} = \frac{1}{ 2 \pi } \; \int_{-1}^{1}  \log\left(  \frac{ 2\sinh( 2 \phi(v) ) \sinh( \hB ) }{ \sinh(  \hB \rho_{\ra}(v) ) } \right)  \frac{ dv }{ \sqrt{1 - v^2 }  },\\
\lim_{z \rightarrow \infty} \log\bh_{\ua}(z)  &=& \log C_{\ua} =  \frac{1}{ 2 \pi } \; \int_{-1}^{1} \log\left( - \frac{ 2 \sinh(  2 \phi(v) ) \sinh( \hmu )}{ \sinh(  \hmu \rho_{\ua}(v) ) } \right)  \frac{ dv }{ \sqrt{1 - v^2 }  }.
\eea
In Section \ref{sec:freeenergy}, we will connect the integrals appearing in (\ref{eq:Cintegral}) to the free energy and establish relation (\ref{eq:asyC}). \\
From (\ref{eq:detr})-(\ref{eq:detu}) and (\ref{eq:resolventP}), one can rewrite the $\wT$ functions in terms of the densities as
\bea\label{eq:mTtowT}
\wT^{\ra}_{1, s} =  \bh_{\ra}^{[+s]} \bh_{\ra}^{[-s]} \mT_{1,s}^{\ra}, \;\;\;\;\;\;\;\wT^{\ua}_{s, 1} =  \bh_{\ua}^{[+s]} \bh_{\ua}^{[-s]} \mT_{s, 1}^{\ua}, \;\;\;\;\;\; s \in \mathbb{N},
\eea
where we have borrowed the notation of \cite{FiNLIE, Cusp2} and denoted
\bea\label{eq:defmT}
{ \mT }^{\ra}_{1,s} = \frac{\sinh\left( \hB(s + \CG_{\ra}^{[s]}- \CG_{\ra}^{[-s]}) \right)}{\sinh( \hB )}, \;\;\;  { \mT }^{\ua }_{s, 1} = \frac{\sinh\left( \hmu(s + \CG_{\ua}^{[s]}- \CG_{\ua}^{[-s]}) \right)}{\sinh( \hmu )}.
\eea
Equation (\ref{eq:mTtowT}) defines a gauge transformation between the $\wT^{\alpha}$ and the $\mT^{\alpha}$ functions, with
\bea
{ \mT }^{\ra }_{0,s} &=& 1, \;\; \;\;{ \mT }^{\ua }_{s,0} = 1, \;\; s \in \mathbb{N},\\
{ \mT }^{\ra }_{2,s} &=& { \mT }^{\ra \, [+s]}_{1,1} \; { \mT }^{\ra \,[-s]}_{1,1},  \;\;{ \mT }^{\ua }_{s, 2} ={ \mT }^{\ua \,[+s]}_{1,1} \; { \mT }^{\ua \,[-s]}_{1,1}, \;\;\;\; s \in \mathbb{N}^+.
\eea
In particular, the $\bh$ factors in (\ref{eq:mTtowT}) cancel in the gauge invariant combinations (\ref{eq:YtoT}).
The gauges $\mT^{\alpha}$ are characterised by the simple large $z$ asymptotics
\bea\label{eq:mTasy}
\mT^{\ra}_{1, s}(z) \sim  \sinh( s \hB )/\sinh( \hB ), \;\;\;\; \mT^{\ua}_{s, 1}(z) \sim  \sinh( s \hmu )/\sinh( \hmu ), \;\;\; s \in\mathbb{N}^+,
\eea
which fixes them uniquely. 
\begin{figure}[t]
\begin{minipage}[b]{0.5\linewidth}
\centering
\includegraphics[width=7.5cm]{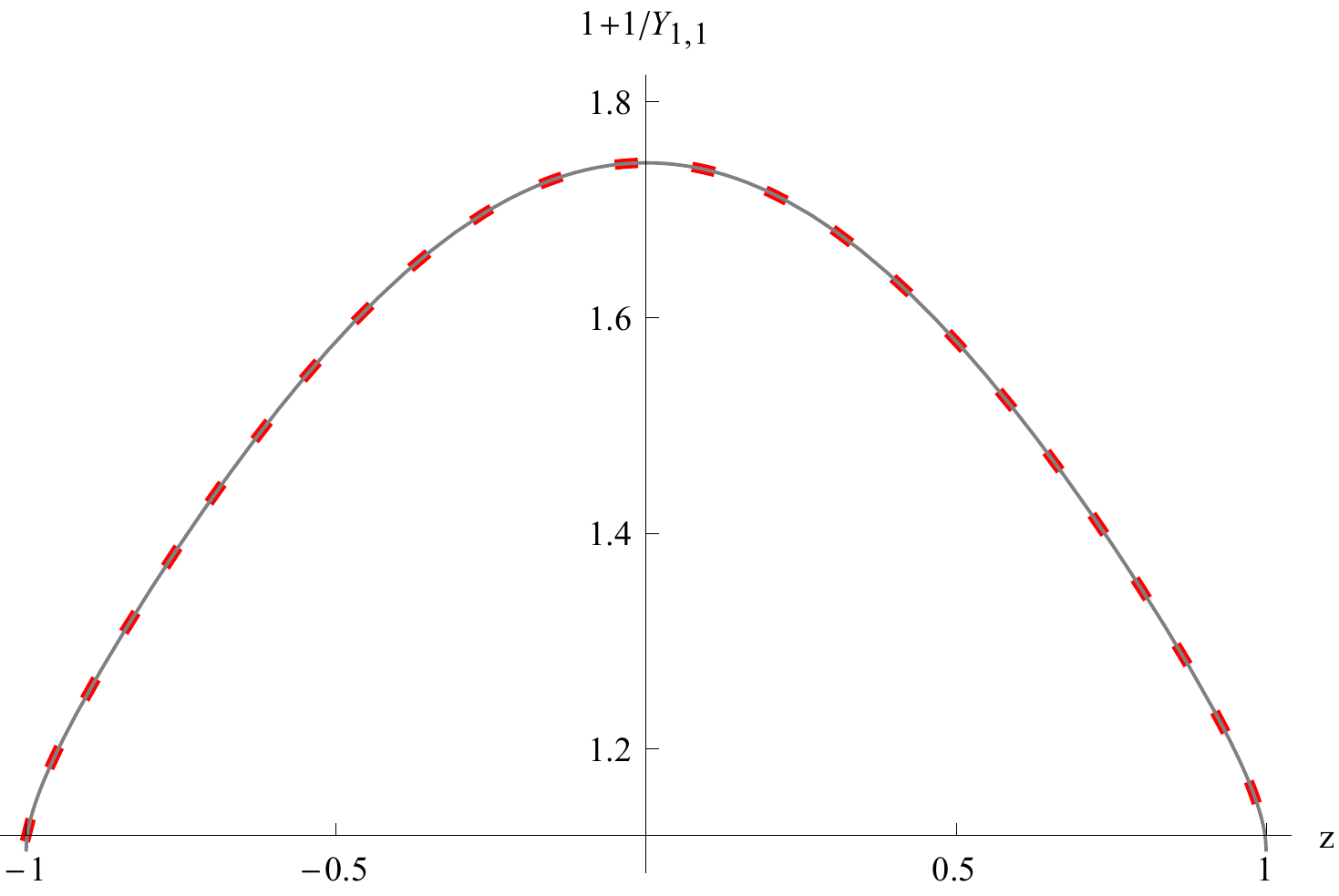}
\end{minipage}
\begin{minipage}[b]{0.5\linewidth}
\centering
\includegraphics[width=7.5cm]{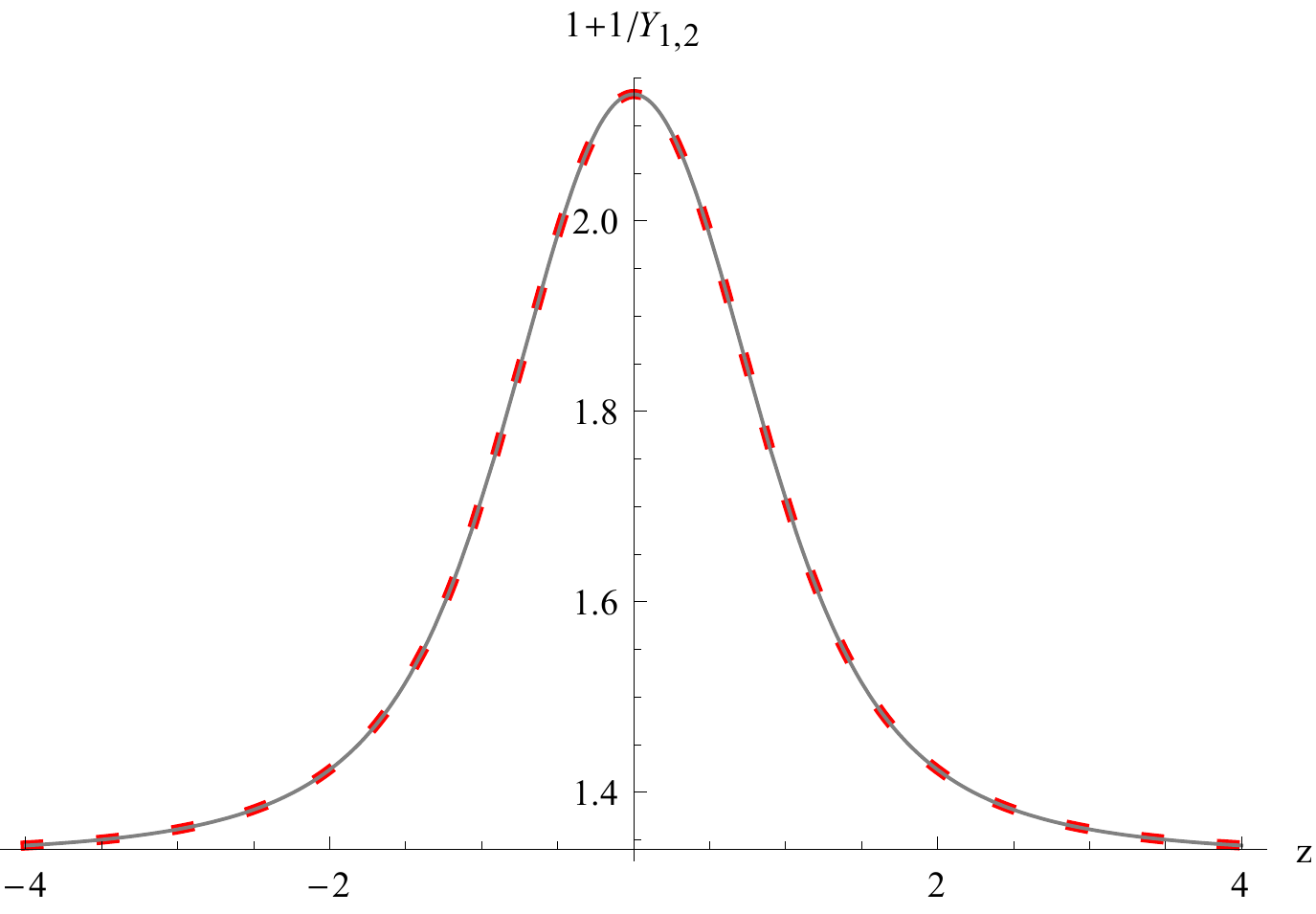}
\end{minipage}
\caption{ Comparison of the numerical solution of the TBA equations (red, dashed line) 
with the new set of nonlinear integral equations (gray, continuous line), at $T=1$, $\bu=1$, $B=0$, $\mu=0$. 
The two curves are interpolations of the numerical solution. The numerical solution of the TBA was obtained by discretising the integrals with a step $\Delta z = 1/60 \simeq 1.7 \times 10^{-2}$, truncating the infinite sums after $60$ terms and cutting off the integrals over the real line  at $z_{\text{max}} = - z_{\text{min}} = 50$. The new NLIEs were solved as explained in Section \ref{sec:numerics}, using $N_{\text{trunc}} = 50$ in (\ref{eq:chebyexpand}). The difference of the curves is of order $10^{-6}$, and is not visible on the scale of the plot. 
\label{fig:TBAcompare}
}
\end{figure}
Finally, in these gauges relations (\ref{eq:T10wr}) become
\bea\label{eq:mT10}
\hspace{-0.3cm}\mT_{1,0}^{\ra}(z) = \frac{\sinh( \hB \rho_{\ra}(z) )}{ 2 \sinh(2\phi(z) ) \sinh( \hB ) } \; e^{-2\phi(z) }, \;\;\;\;
\mT_{0,1}^{\ua}(z) = -\frac{\sinh( \hmu \rho_{\ua}(z) )}{ 2 \sinh(2\phi(z) ) \sinh( \hmu) } \; e^{2\phi(z) }.
\eea
As discussed in Section \ref{sec:numerics}, the densities can be computed by solving numerically a set of NLIEs, which can be derived from the constraint (\ref{eq:wT}). 
Some examples of solutions are displayed in Figures \ref{rho1} and \ref{rho2}. 
Finally, let us remark again that the information contained in the pair $\left\{ \rho_{\ra},\rho_{\ua} \right\}$ 
is fully equivalent to the knowledge of the solution of the TBA equations. In Figure \ref{fig:TBAcompare}, the TBA solution 
is compared with the same quantity reconstructed from the parametrisation (\ref{eq:YtoT}),(\ref{eq:defmT}).
\subsection{The free energy}\label{sec:freeenergy}
The purpose of this Section is to derive a simple formula for the free energy in terms of the densities. The starting point is equation  (\ref{eq:energyrw}), written in the equivalent form: 
 \bea\label{eq:ETtelesc}
&& f/T + \hmu = - \bu/T-\log(2 \cosh( \hB )) -\frac{1}{2\pi}\int_{-1}^{1}\log\left(( 1 + 1/Y_{1, 1}(z) )( 1 + Y_{2,2}(z) ) \right) \thh'(z) \, dz \nn\\
&&-\frac{ 1}{2\pi } \sum_{n=1}^{\infty} \InR \; \log\left( \frac{ \bar{\mT}^{\ra}_{1,n+1}(z+ i \bu)  \bar{\mT}^{\ra}_{1,n+1}(z-i \bu)}{  {\bar \mT}^{\ra}_{1,n}(z)  {\bar \mT}^{\ra}_{1,n+2}(z) }\right) \, \thh'_n(z) \; dz,
\eea
where we have set ${\bar \mT }^{\ra}_{1,s} = {\mT }^{\ra}_{1,s} \sinh(\hB )/(\sinh( s \hB) )$, so that $\ln {\bar \mT }^{\ra}_{1,s}(z) \sim 0$ at large $z$. 
Adopting the ``telescoping'' technique used in \cite{BalogHegedus, FiNLIE, GromovSever}, it is now possible to remove the infinite sum appearing on the rhs of (\ref{eq:ETtelesc}). 
Since ${\bar \mT }^{\ra}_{1,n}$ 
is analytic for $|\IIm(z)| < (n+1) \bu$, we can shift the integration contours and prove that, for $n \in \mathbb{N}^+$,
\bea
&&\InR \; \log\left(   \bar{\mT}^{\ra , \left[+1\right]}_{1,n+1}(z)  \bar{\mT}^{\ra, \left[-1\right]}_{1,n+1}(z) \right)\, \thh'_n(z) \, dz =
\InR \; \log\left(  \bar{\mT}^{\ra}_{1,n+1}(z)\right) \; \left( \thh'_{n+1}(z) + \thh'_{n-1}(z)
 \right) \, dz. \nn\\
\eea
This shows that many cancellations take place in (\ref{eq:ETtelesc}), leading to
\bea\label{eq:ET2}
f/T + \hmu &=& -\bu/T-\frac{1}{2 \pi}\int_{-1}^{1}\log\left(( 1 + 1/Y_{1, 1}(z) )( 1 + Y_{2,2}(z) ) \right) \, \thh'(z) \, dz \\
&& - \frac{1}{\pi } \int_{-1}^{1} \log\mT^{\ra}_{1,2}(z) \, \thh'(z) \, dz + \frac{ 1}{2\pi } \InR \; \log\mT^{\ra}_{1,1}(z) \, \thh'_1(z) \, dz.\nn
\eea
Using (\ref{eq:YtoT}) and the fact that $Y_{2,2} = 1/{ \widetilde Y }_{1,1}$, (\ref{eq:ET2}) becomes
\bea\label{eq:ET3}
f/T + \hmu &=& - \bu/T - \frac{1}{2\pi } \int_{-1}^{1} \log\left(\frac{ \mT^{\ra \,[+1]}_{1,1}(z) \, \mT^{\ra \,[-1]}_{1,1}(z) \widetilde{\mT^{\ra \,[+1]}_{1,1}}(z) \widetilde{\mT^{\ra \,[-1]}_{1,1}}(z) }{ \mT_{1,0}^{\ra}(z) \,  \widetilde{\mT_{1,0}^{\ra}}(z)  }\right) \, \thh'(z) \, dz \nn \\
&+&\frac{1}{2\pi } \InR \;  \log\mT^{\ra}_{1,1}(z) \,  \thh'_1(z) \, dz.
\eea  
The convolution appearing in the first line of (\ref{eq:ET3}) can be viewed as a contour integral around the cut of $\thh'(z)$, and, deforming this contour to a pair of infinite lines, one can write
\bea
&&\int_{-1}^{1} \log\left(  { \mT^{\ra\,[+1]}_{1,1} }(z) {\mT^{\ra\,[-1]}_{1,1} }(z) \widetilde{{\mT^{\ra\,[+1]}_{1,1}}}(z) \widetilde{{\mT^{\ra\,[-1]}_{1,1}}}(z) \right) \, \thh'(z) \, dz \nn \\
&=& \left( \int_{\R + i 0^+} -  \int_{\R - i 0^+} \right) \log\left( { \mT^{\ra\,[+1]}_{1,1} }(z) {\mT^{\ra\,[-1]}_{1,1} }(z) \right) \, \thh'(z) \, dz.
\eea
Shifting the line contours appearing in the latter expression (and pushing some of them to infinity), one can show that  the integrals involving $\mT_{1,1}^{\ra}$ in equation (\ref{eq:ET3}) actually cancel each other out.
Substituting (\ref{eq:mT10}) for $\mT_{1,0}^{\ra}$, we finally arrive at the desired result:
\bea\label{eq:ET4}
f + \mu &=& - \bu - \frac{T}{2\pi } \int_{-1}^{1} \log\left(\frac{ 4 \, \sinh^2( 2 \phi(z) ) \sinh^2(\hB)}{ \sinh^2( \hB \rho_{\ra}(z) )}\right) \; \frac{dz}{ \sqrt{ 1 - z^2 } }.
\eea
Starting from the alternative expression (\ref{eq:energyuw}) for the free energy, working in the gauge $\mT^{\ua}$ and following the same steps, we also obtained
\bea\label{eq:ET5}
f + \mu&=& \bu - \frac{T}{2\pi } \int_{-1}^{1} \log\left(\frac{ 4 \, \sinh^2( 2 \phi(z) ) \sinh^2(\hmu) }{ \sinh^2(\hmu \rho_{\ua}(z) )}\right) \; \frac{dz}{ \sqrt{ 1 - z^2 } }.
\eea
The comparison between (\ref{eq:ET4}), (\ref{eq:ET5}) and (\ref{eq:Cintegral}) finally gives the formula 
(\ref{eq:asyC}), showing that the free energy $f$ appears in the asymptotics of the $\bP$'s. 
\subsection{Energy-carrying Bethe roots}\label{sec:energyBethe}
Let us look more closely at the kernels appearing in (\ref{eq:ET4}) and (\ref{eq:ET5}). From (\ref{eq:exprho0}), we have
\bea\label{eq:sinhrho}
\frac{ \sinh^2( 2 \phi(z) )  }{ \sinh^2( \hB \rho_{\ra}(z) )} = \tP_{\p}^{\ra}(z) \bP_{\p}^{\ra}(z)  \tP_{\m}^{\ra}(z) \bP_{\m}^{\ra}(z),
\eea
where we have used (\ref{eq:constraint31}). The same relation with $\hB \rightarrow \hmu$, $\bigra \rightarrow \bigua$ is valid for the density $\rho_{\ua}$. Thus, the kernels entering the energy formula have logarithmic singularities in correspondence to the zeros of the $\bP$ functions. 
Should any of these zeros cross the integration contours in  (\ref{eq:ET4}),(\ref{eq:ET5}), the energy would pick extra residue terms of the form $ i T \, {\widetilde \thh}(z_i) $, 
as is easy to verify by first performing an integration by parts\footnote{One could also express the ground state free energy as an infinite sum $ i T \sum_j \thh(z_j) $ (which should be regularised).}. 
This suggests that zeros of the $\bP$ functions can be regarded as energy-carrying Bethe roots for the finite temperature spectrum. 
According to our assumptions, for the ground state with real $B$ and $\mu$ all the zeros lie on the second sheet, but they may cross the interval and move to the first sheet under analytic continuation in $B$ and $\mu$. We will return on this point in Section \ref{sec:freefermion} below. 
\section{Exact Bethe Ansatz and relation with the Quantum Transfer Matrix}
\label{sec:QTM}
In this Section we will recast the closed set of conditions (\ref{eq:constraint31})-(\ref{eq:wT}) into another, interesting form, which reveals the presence of an exact Bethe Ansatz. This second formulation is perhaps closer in spirit to the $\bP \mu$-system  
of AdS/CFT~\cite{QSC}, which was one of the main sources of inspiration for the present work.  
\subsection{Formulation as a coupled Riemann-Hilbert problem}
We start by considering the analytic continuation of (\ref{eq:wT}) through the cuts at $z \in (-1,1) \pm i \bu$, which yields 
\bea\label{eq:fracT}
\frac{ { \bP}_{\p}^{\ra \, [\pm 2]}(z) \, { \widetilde \bP }_{\m}^{\ra}(z) - { \bP}_{\m}^{\ra \,[\pm2]}(z) \, { \widetilde \bP }_{\p}^{\ra}(z) }{ { \bP}_{\p}^{\ua \,[\pm2]}(z) \, { \widetilde \bP }_{\m}^{\ua}(z) - {\bP}_{\m}^{\ua \,[\pm2]}(z) \, { \widetilde \bP }_{\p}^{\ua}(z) } = e^{ \mp  ( \phi^{[\pm 2]}(z) +  \phi(z) )  }.
\eea
Two of the ${ \widetilde \bP }$ functions in (\ref{eq:fracT}) can be eliminated using the constraints (\ref{eq:constraint31}),  (\ref{eq:constraint32}). Solving (\ref{eq:fracT}) for the remaining two ${ \widetilde \bP }$'s, we find
\bea\label{eq:new}
{ \widetilde \bP }_a^{\ra}(z) \bP^{\ua}_b(z) &=& \bF_{ab}^{[+1]}(z) e^{\phi(z)} +  \bF_{ab}^{[-1]}(z) e^{-\phi(z)}, \; \;\;\;\; a, b = \p, \m,
\eea
where 
$\bF_{ab}$ has a simple explicit expression: 
\bea\label{eq:defF}
\bF_{ab}(z) &=& \frac{ e^{\phi^{[-1]}(z)} \bP_a^{\ra \,[+1]}(z) \bP_b^{\ua \,[-1]}(z) + e^{-\phi^{[+1]}(z)} \bP_a^{\ra \, [-1]}(z)  \bP_b^{\ua \, [+1]}(z) }{ \wT_{1,1}^{\ra}(z) }, \; a, b = \p,\m.
\eea
Relation (\ref{eq:defF}) shows that $\bF_{ab}(z)$ could -- in principle -- have a pair of branch cuts on the lines $\IIm(z) = \pm i \bu$. 
However, by shifting equation (\ref{eq:new}) of $\pm i \bu$, and evaluating the discontinuity across the branch cut, we find
\bea
e^{-\phi^{[\pm1]}} \, \disc\left[ \bF_{ab}^{[\pm1]}  \right]
= \disc\left[ { \widetilde \bP }_a^{\ra \,[\pm1]} \, \bP^{\ua \,[\pm 1]}_b  \right] = 0, 
\eea
which implies that $\bF_{ab}$ is analytic in the whole complex plane. The absence of branch cuts can also be directly verified by using the definition (\ref{eq:defF}): the quantities $\bF_{ab}^{[\pm1]}  - { \widetilde  \bF _{ab}^{[\pm1]} }$ vanish since they are proportional to the difference of the lhs and rhs of (\ref{eq:wT}). 
Notice that the lhs of (\ref{eq:new}) has no poles. As a consequence, it is possible to prove that the $\bF$ functions, appearing on the rhs, are entire on the whole complex plane. This can be also deduced from  equation (\ref{eq:Fshift}) below.

These newly introduced quantities fulfill, together with the $\bP$'s, a closed set of functional relations. The fundamental set of equations is
\bea
\bP_{\p}^{\ra}(z) { \widetilde \bP}^{\ra}_{\m}(z) - { \widetilde \bP}^{\ra}_{\p}(z) \bP^{\ra}_{\m}(z) &=&   -2 \;\sinh( 2 \phi(z) ),\label{eq:closed2} \\
\bP^{\ua}_{\p}(z) { \widetilde \bP}^{\ua}_{\m}(z) - { \widetilde \bP}^{\ua}_{\p}(z) \bP^{\ua}_{\m}(z) &=& 2  \; \sinh( 2 \phi(z) ),\label{eq:closed3}\\
{ \widetilde \bP }_a^{\ra}(z) \bP^{\ua}_b(z) = \bF_{ab}^{[+1]}(z) e^{\phi(z)} &+&  \bF_{ab}^{[-1]}(z) e^{-\phi(z)}, \;\;\;a, b = \p, \m,\label{eq:closed4}
\eea
with 
the requirement that the $\bF$ functions have no cuts and no poles, and the $\bP$'s have only one cut on each of their two sheets, and no poles. 
These equations make explicit the nonlinear Riemann-Hilbert nature of the problem, since they show how the two branches of the $\bP$'s are connected through the entire $2 \times 2$ matrix $\bF_{ab}$. In this respect, they are reminiscent of the $\bP\mu$-system.
It is important to underline that (\ref{eq:closed2})-(\ref{eq:closed4}) are completely equivalent to (\ref{eq:constraint31})-(\ref{eq:wT}). 
In particular, it is possible to show that the definition (\ref{eq:defF}) and the relation (\ref{eq:wT}) are both hidden in these equations. 
 
Equations (\ref{eq:closed2})-(\ref{eq:closed4}) imply the existence of many other functional relations among the $\bF$'s and the $\bP$'s. For instance:
\bea
\bF_{\p\p}(z) \, \bF_{\m\m}(z) - \bF_{\p\m}(z) \, \bF_{\m\p}(z) &=& -1 \\
\bF_{\p\p}^{[+1]}(z ) \, \bF_{\m\m}^{[-1]}(z) + \bF_{\m\m}^{[+1]}(z) \, \bF_{\p\p}^{[-1]}(z) &=& \bF_{\p\m}^{[+1]}(z ) \, \bF_{\m\p}^{[-1]}(z) +\bF_{\m\p}^{[+1]}(z) \, \bF_{\p\m}^{[-1]}(z)  + 2 \cosh( 2 \phi(z) ). \nn
\eea
Finally, as already  remarked in Section \ref{sec:intro},   the Hubbard Hamiltonian (\ref{H})  has, for $B=\mu=0$, a hidden 
$SO(4) \cong SU(2) \times SU(2)/\ZZ_2$  symmetry  \cite{HubbardSO4}. 
While this symmetry was not very evident in the original 
Y-system and discontinuity relations, it appears to be nicely 
encoded in the structure of equations (\ref{eq:closed2})-(\ref{eq:closed4}). 
At generic values of the magnetic field or chemical potential, the symmetry is  broken by the boundary conditions (\ref{eq:Pasy}).
\subsection{The exact Bethe Ansatz}
The relations obtained in the previous Section contain an exact Bethe Ansatz constraining the position of the zeros of the $\bP$  and $\bF$ functions. 
Starting from (\ref{eq:new}), we immediately get 
\bea\label{eq:Fshift}
\bF_{ab}^{[\pm1]}(z) = \pm \frac{1}{2} \frac{  e^{\pm\phi(z) } \, { \widetilde \bP }_a^{\ra}(z) \bP_b^{\ua}(z) -  e^{\mp\phi(z) } \bP_a^{\ra}(z) { \widetilde \bP }_b^{\ua}(z) }{\sinh( 2 \phi(z) ) }, \;\;\;\;a, b=\p, \m.
\eea
Combining these expressions to form bilinear combinations of the $\bF_{a b}$'s reveals a further set of interesting  relations of quantum Wronskian type. For instance, using (\ref{eq:Fshift}) and (\ref{eq:constraint31}),(\ref{eq:constraint32}), we obtain
\bea\label{eq:eq10}
\bF_{\p\p}^{[+1]}(z) \bF_{\p\m}^{[-1]}(z) - \bF_{\p\p}^{[-1]}(z) \bF_{\p\m}^{[+1]}(z) = { \bP_{\p}^{\ra} }(z) { \widetilde \bP }_{\p}^{\ra}(z).
\eea
The rhs of (\ref{eq:eq10}) defines an entire function on the whole complex plane. 
It is convenient to isolate the exponential prefactors and set
\bea\label{eq:Q1}
\mathcal{Q}_{\p}^{\ra}(z) = e^{ i \hB z/\bu } \, \bP_{\p}^{\ra}(z) \, { \widetilde \bP }_{\p}^{\ra}(z), \;\;\;\; \f_{\p\m}(z) =  e^{ \frac{i}{2} (\hB - \hmu ) z / \bu  } \,\bF_{\p\m}(z). 
\eea
The $a=\p$, $b=\m$ case of equation (\ref{eq:new}) can then be rewritten as
\bea\label{eq:eqb}
\f_{\p\m}^{[+1]}(z) e^{\hB - \hmu +\phi(z) } + 
\f_{\p\m}^{[-1]}(z) e^{-\hB+\hmu -\phi(z)}  = { \widetilde \bP }_{\p}^{\ra}(z) \bP^{\ua}_{\m}(z) e^{\frac{i}{2}(\hB - \hmu ) z / \bu }.
\eea
We assume that for the ground state all zeros of the $\bP$ functions are on the second sheet; therefore the zeros of the rhs of (\ref{eq:eqb}) coincide with the zeros of $\mQ_{\p}^{\ra}$, 
giving the first of the Bethe Ansatz equations quoted in Section \ref{sec:intro}. 
The second BA equation can be derived by taking ratios of the expressions obtained by evaluating (\ref{eq:eqb}) at zeros of $\f_{{\p\m}}^{[\pm]}$. 
The resulting set of quantisation conditions is:
\bea
e^{ \hB - \hmu }  \; \frac{ \f_{{\p\m}}(s_i + i \bu )}{ \f_{\p\m}(s_i - i \bu ) } &=& - e^{ -2 \phi( s_i ) }, \;\;\; \text{at}\;\; \mathcal{ Q}_{\p}^{\ra}(s_i) = 0, \label{eq:BA3}\\
e^{ -2 \hmu  } \;  \frac{ \f_{\p\m}( w_{\alpha} + 2 i \bu ) }{ \f_{\p\m}( w_{\alpha} - 2 i \bu ) } &=& -\frac{ \mathcal{ Q}_{\p}^{\ra}( w_{\alpha} + i \bu) }{\mathcal{ Q}_{\p}^{\ra}( w_{\alpha} - i \bu) }, \;\;\; \text{at} \;\; \f_{\p\m}( w_{\alpha} ) = 0.\label{eq:BA4}
\eea
Very interestingly, these equations appear to be the infinite Trotter number limit of the BA diagonalising the Quantum Transfer Matrix of J\"uttner, Kl\"umper and Suzuki~\cite{DDVHubbard}. 
We remind the reader that the QTM is defined as a discrete object acting on a $\NT$-site quantum space ($\NT$ is known as the Trotter number), and that the thermodynamics of the Hubbard model is described by the largest eigenvalue of the QTM in the limit $\NT \rightarrow \infty$. For the ground state at finite even values of $\NT$, the BA of \cite{DDVHubbard} is 
\bea
e^{ \hB - \hmu }  \; \frac{ q_{2}( s_i + i \bu )}{ q_2(s_i - i \bu ) } &=& -b_{\NT}(s_i), \;\;\; i=1, \dots, \NT, \label{eq:BAN}\\
e^{ -2 \hmu  } \;  \frac{ q_2( w_{\alpha} + 2 i \bu ) }{ q_2( w_{\alpha} - 2 i \bu ) } &=& -\frac{ q_1( w_{\alpha} + i \bu) }{q_1( w_{\alpha} - i \bu) }, \;\;\;\alpha = 1, \dots, \NT/2. \label{eq:BAN2}
\eea
where\footnote{See Appendix \ref{app:dictionary} for more details. }  $\lim_{\NT\rightarrow \infty} b_{\NT}(z) = e^{-2\phi(z)}$ and 
\bea
q_1(z) = \prod_{i=1}^{\NT} \left(z - s_i \right), \;\;\;\; q_2(z) = \prod_{\alpha=1}^{\NT/2} \left(z - w_{\alpha} \right).\label{eq:polynomials}
\eea
 The functions appearing in the BA equations (\ref{eq:BA3}),(\ref{eq:BA4}) are the continuum version of these polynomials. 
 In fact, we believe that they can be factorised over their zeros as infinite products of the form\footnote{
We expect that the $\bP$ functions can be factorised in the following form (cf \cite{Cusp2}), 
\bea
(2 \sinh( \hB) )^{\frac{1}{2} } \, e^{ \bpm \frac{i}{2} \hB  z/\bu   } \,{ \bP }_{\bpm}^{\ra}(z) &=&  C_{\ra} \, \prod_{i=1}^{\infty} \left(1 + 1/( z_i^{(\ra, \bpm)}\; \x(z)  ) \right)  \left(1 + 1/( z_{-i}^{(\ra, \bpm)} \; \x(z) ) \right),
\eea
where $z_{i}^{(\ra, \bpm)} = \x( \, s_i^{(\ra, \bpm)} \, )$,  
and the Zhukovsky map $\x(z)$ is defined in (\ref{eq:Zhukovsky}). Again, the pairing of zeros is to guarantee the convergence of the product.
} 
\bea\label{eq:Hadamard}
\frac{\f_{\p\m}(z) }{ \f_{\p\m}(0)} =  \prod_{\alpha=1}^{\infty} \left(1 -\frac{z}{w_{\alpha}^{(\p\m)}}\right) \left(1 -\frac{z}{w_{-\alpha}^{(\p\m)}}\right), \;\;\;\;  \frac{\mathcal{Q}_{\p}^{\ra}(z) }{ \mathcal{Q}_{\p}^{\ra}(0)}  =  \prod_{i =1 }^{\infty} \left(1 -\frac{z}{s_i^{(\ra, \p)}}\right) \left(1 -\frac{z}{s_{-i}^{(\ra, \p)}}\right), \nn\\
\eea
where $\RRe( s_{-i} ) = - \RRe( s_{i} )$, $\RRe( w_{-\alpha} ) = - \RRe( w_{\alpha} )$. In (\ref{eq:Hadamard}), the zeros have been paired up in order to make the product convergent. This rearrangement is necessary, since the zeros accumulate at infinity with an evenly spaced asymptotic distribution
(see Sections \ref{sec:freefermion} and \ref{sec:numerics}). The regularisation (\ref{eq:Hadamard}),  without extra Hadamard factors, is consistent with the infinite Trotter number limit of (\ref{eq:BAN})-(\ref{eq:polynomials}). 

In addition to (\ref{eq:BA3}), (\ref{eq:BA4}), there are naturally other equivalent sets of BA equations.   For completeness, let us discuss the general structure. From (\ref{eq:Fshift}) and (\ref{eq:constraint31}), (\ref{eq:constraint32}), one can obtain the following quantum Wronskian-type relations
\bea\label{eq:wronskians}
\bF_{\bpm \bpm}^{[+1]}(z) \bF_{\bpm \bmp}^{[-1]}(z) - \bF_{\bpm \bpm}^{[-1]}(z) \bF_{\bpm \bmp}^{[+1]}(z) &=& \pm \bP_{\bpm}^{\ra}(z) { \widetilde \bP }_{\bpm}^{\ra}(z), \label{eq:wronskiansfirst}\\
\bF_{\bpm \bpm}^{[+1]}(z) \bF_{\bmp\bpm}^{[-1]}(z) - \bF_{\bpm\bpm}^{[-1]}(z) \bF_{\bmp\bpm}^{[+1]}(z) &=& \pm \bP_{\bpm}^{\ua}(z) { \widetilde \bP }_{\bpm}^{\ua}(z), \\
\bF_{\p\p}^{[+1]}(z) \bF_{\m\m}^{[-1]}(z) - \bF_{\p\p}^{[-1]}(z) \bF_{\m\m}^{[+1]}(z) &=& \bP_{\p}^{\ua}(z) { \widetilde \bP }_{\m}^{\ua}(z) +  \bP_{\p}^{\ra}(z) { \widetilde \bP }_{\m}^{\ra}(z), \\
\bF_{\p\m}^{[+1]}(z) \bF_{\m\p}^{[-1]}(z) - \bF_{\p\m}^{[-1]}(z) \bF_{\m\p}^{[+1]}(z) &=& \bP_{\p}^{\ua}(z) { \widetilde \bP }_{\m}^{\ua}(z) - \bP_{\m}^{\ra}(z) { \widetilde \bP }_{\p}^{\ra}(z), \label{eq:wronskianslast}
\eea
which allow one to derive alternative pairs of BA equations. 
Generalising (\ref{eq:Q1}), we define
\bea\label{eq:otherQ}
\mQ_{a}^{\ra}(z) &=& e^{  i \hB_a z/\bu } \, \bP_{a}^{\ra}(z) \, { \widetilde \bP }_{a}^{\ra}(z), \label{eq:otherQ1}\\
\mQ_{a}^{\ua}(z) &=& e^{  i \hmu_a z/\bu } \, \bP_{a}^{\ua}(z) \, { \widetilde \bP }_{a}^{\ua}(z), \label{eq:otherQ2}  \\
\f_{ab}(z) &=&  e^{ \frac{i}{2} ( \hB_a + \hmu_b ) z / \bu  } \,\bF_{ab}(z),  \;\;\;\;\;\;\; a, b = \p, \m, \label{eq:otherQlast}
\eea
where 
$\hB_{\pm} = \pm \hB$, $\hmu_{\pm} = \pm \hmu$. We expect that all the $\f_{ab}$ and $\mQ_a$ functions thus introduced admit a factorisation of the form (\ref{eq:Hadamard}), 
each with a different set of zeros. The systems of BA equations fulfilled by these functions 
can be easily obtained from 
(\ref{eq:BA3}) by symmetry. 
We can schematically summarise these symmetries as
\bea
\mQ_{\m}^{\ra}(z ; \hB, \hmu ) &=& \mQ_{\p}^{\ra}(z ; -\hB, \hmu ), \;\;\;\;\; \mQ_{\m}^{\ua}(z ; \hB, \hmu ) = \mQ_{\p}^{\ua}(z ; \hB, -\hmu ),\\
\f_{\m a}(z ; \hB, \hmu ) &=& \f_{\p a}(z ; -\hB, \hmu ), \;\;\f_{a\m}(z ; \hB, \hmu ) = \f_{a\p}(z ; \hB, -\hmu ), 
\eea
with $a = \p, \m$. 
In addition, we have the substitution rule 
\bea
\mQ_a^{\ra} \lra \mQ_a^{\ua},\;\;\;\; \phi \lra { \tilde \phi } = -\phi, \;\;\;\; a=\p, \m.
\eea
Finally, let us point out that the definitions (\ref{eq:otherQ1}), (\ref{eq:otherQ2}) can be inverted as
\bea
\log \bP_{a}^{\ra}(z) &=& - \frac{i}{2} \hB_a z/\bu  + \frac{\sqrt{ 1 - z^2 }}{ 2 \pi i } \; \int_{-1}^{1} \log{  \mQ^{\ra}_{a}(v)  } \,  \frac{ dv }{ \sqrt{1 - v^2 } \; ( v - z ) },\\
\log \bP_{a}^{\ua}(z) &=& -\frac{i}{2} \hmu_a z/\bu  + \frac{\sqrt{ 1 - z^2 }}{ 2 \pi i } \; \int_{-1}^{1} \log{  \mQ^{\ua}_{a}(v)  } \,  \frac{ dv }{ \sqrt{1 - v^2 } \; ( v - z ) },
\eea
($a = \p, \m$), leading to the following equivalent formulae for the free energy:
\bea
f + \mu  &=& -\bu  -\frac{T}{ \pi } \; \int_{-1}^{1} \log( 2\sinh( \hB ) \, \mQ^{\ra}_{\bpm}(v) ) \,  \frac{ dv }{ \sqrt{1 - v^2 } } =  \bu  -\frac{T}{ \pi } \; \int_{-1}^{1} \log( 2 \sinh( \hmu ) \,  \mQ^{\ua}_{\bpm}(v) ) \,  \frac{ dv }{ \sqrt{1 - v^2 } }.\nn\\
\eea

\subsection{Relation with the Quantum Transfer Matrix}
By comparison with the results of \cite{DDVHubbard} we can uncover the physical meaning of $\wT_{1,1}^{\ra}$, $\wT_{1,1}^{\ua}$ and show that these functions are simply related to the eigenvalues  of the Quantum Transfer Matrix. Starting from (\ref{eq:new}), we get
\bea
\left(  e^{-2 \phi^{[+1]}(z) } + \frac{ \bF_{ab}^{[+2]}(z)  }{ \bF_{ab}(z)  }  \right) &=&e^{ -\phi^{[+1]}(z)  } \, \frac{{ \widetilde \bP }_a^{\ra \,[+1]}(z) \, \bP^{\ua \,[+1]}_b(z) }{ \bF_{ab}(z)  }, \label{eq:firsteq} \\
\left( e^{ 2  \phi^{[-1]}(z) } + \frac{ \bF_{ab}^{[-2]}(z)  }{ \bF_{ab}(z)  }  \right) &=&e^{ \phi^{[-1]}(z)  } \, \frac{{ \widetilde \bP }_a^{\ra \,[-1]}(z) \, \bP^{\ua \,[-1]}_b(z) }{ \bF_{ab}(z)  },  \;\;\;a, b = \p, \m.  \label{eq:secondeq}
\eea
Multiplying (\ref{eq:firsteq}) by $\bP^{\ra\,[-1]}_a /  { \widetilde \bP }^{\ra \,[+1]}_a $ and (\ref{eq:secondeq}) by $\bP^{\ra\,[+1]}_a / { \widetilde \bP }^{\ra \,[-1]}_a $ and adding them, we find, after using (\ref{eq:defF})
\bea
&&\frac{ \bP^{\ra \,[-1]}_a(z) }{  { \widetilde \bP }^{\ra \,[+1]}_a(z) }  \, \left(  e^{-2 \phi^{[+1]}(z) } + \frac{ \bF_{ab}^{[+2]}(z) }{ \bF_{ab}(z) }  \right) + \frac{\bP^{\ra \,[+1] }_a(z) }{ { \widetilde \bP }^{\ra \,[-1]}_a(z) } \,\left(  e^{ 2 \phi^{[-1]}(z) } + \frac{ \bF_{ab}^{[-2]}(z) }{ \bF_{ab}(z) }  \right) = \label{eq:idenprecise}\\
&&\frac{ \bP^{\ra \,[-1]}_a(z) \, \bP_b^{\ua\,[+1]}(z) \,e^{- \phi^{[+1]}(z) }   + \bP^{\ra \,[+1]}_a(z)  \, \bP^{\ua \,[-1]}_b(z) \,e^{ \phi^{[-1]}(z) }  }{ \bF_{ab}(z) } =  \wT_{1,1}^{\ra}(z) \label{eq:idenprecise2}.
\eea
The combination in (\ref{eq:idenprecise}) agrees with the form of the Quantum Transfer Matrix eigenvalues (see equation (15) in \cite{DDVHubbard}), namely 
\bea
(\Lambda )_{\text{ref \cite{DDVHubbard}}} \leftrightarrow \wT_{1,1}^{\ra}.
\eea
The precise details  of this identification are given in Appendix \ref{app:dictionary}. Notice that, as a consequence of the absence of poles for the $\bP$ functions, this quantity does not have poles on any sheet, and that this pole-free condition gives precisely the Bethe Ansatz. 
Furthermore, setting ${ \widetilde \bP }_a^{\ra} \bP_a^{\ra} = { \bar \mQ }_a^{\ra}$,  we can write
\bea
\wT_{1,1}^{\ra} = \bP^{\ra \,[+1]}_a \, \bP^{\ra \,[-1] }_a \left[ \frac{1 }{\bar{\mathcal{ Q }}_a^{\ra[+1]} }  \,\left(  e^{-2 \phi^{[+1]} } + \frac{ \bF_{ab}^{[+2]} }{ \bF_{ab} }  \right) + \frac{1}{\bar{\mathcal{Q} }_a^{\ra[-1]}}  \,\left(  e^{2 \phi^{[-1]} } + \frac{ \bF_{ab}^{[-2]} }{ \bF_{ab} }  \right) \right].  
\eea
In the quantity in the square brackets we recognise the infinite Trotter number limit of the ``auxiliary'' transfer matrix eigenvalues $\Lambda^{\text{aux}}$ introduced in \cite{DDVHubbard}, equation (25). 
Setting $a=\p$, the identification is
\bea
(\Lambda^{\text{aux} } )_{\text{ref \cite{DDVHubbard} }} \leftrightarrow   \frac{  \wT_{1,1}^{\ra}  }{ \bP^{\ra \,[+1]}_{\p} \,  \bP^{\ra \,[-1]}_{\p}  }.
\eea
\section{The free fermion limit}
\label{sec:freefermion}
The exact solution of the TBA equations at $\bu=0$ was found already by Takahashi in \cite{Takahashi}. 
It is interesting to recover this result starting from relations (\ref{eq:constraint31})-(\ref{eq:wT}). In the limit $\bu \rightarrow 0^+$, the shifts in the $\bP$ functions shrink to zero and, for $\RRe(z) \in (-1, 1)$, the $\bP$ functions collapse to their values above/below the cut. Therefore, 
\bea
\mathbb{ T }_{1,1}^{ \ra } &\sim& e^{\hB} \bP_{\p}^{\ra} \, { \widetilde \bP }_{\m}^{\ra} - e^{-\hB} \bP_{\m}^{\ra} \, { \widetilde \bP }_{\p}^{\ra}, \\
\mathbb{ T }_{1,1}^{ \ua } &\sim& e^{\hmu} \bP_{\p}^{\ua} \, { \widetilde \bP }_{\m}^{\ua} - e^{-\hmu } \bP_{\m}^{\ua} \, { \widetilde \bP }_{\p}^{\ua}, 
\eea
and equation (\ref{eq:wT}) becomes
\bea\label{eq:constrfree1}
e^{\hB} \bP_{\p}^{\ra} \, { \widetilde \bP }_{\m}^{\ra} - e^{-\hB} \bP_{\m}^{\ra} \, { \widetilde \bP }_{\p}^{\ra} = \left( e^{\hmu} \bP_{\p}^{\ua} \, { \widetilde \bP }_{\m}^{\ua} - e^{-\hmu } \bP_{\m}^{\ua} \, { \widetilde \bP }_{\p}^{\ua} \right) \, e^{ -2 \phi }.
\eea
A second constraint is simply obtained by continuing (\ref{eq:constrfree1}) to the second branch:
\bea\label{eq:constrfree2}
e^{\hB} \tP_{\p}^{\ra} \, \bP_{\m}^{\ra} - e^{-\hB} \tP_{\m}^{\ra} \, \bP_{\p}^{\ra} = \left( e^{\hmu} \tP_{\p}^{\ua} \, \bP_{\m}^{\ua} - e^{-\hmu } \tP_{\m}^{\ua} \, \bP_{\p}^{\ua} \right) \, e^{ 2 \phi }.
\eea
Solving (\ref{eq:constrfree1}),(\ref{eq:constrfree2}) with the aid of (\ref{eq:constraint31}),(\ref{eq:constraint32}), we find 
\bea\label{eq:explP}
\sinh(\hB) \, \bP_{\p}^{\ra}(z) { \widetilde \bP}_{\m}^{\ra}(z) = 2\cosh( \phi(z) - (\hB + \hmu )/2 )  \; \cosh(\phi(z) - (\hB - \hmu )/2 ), \\
\sinh(\hmu) \, \bP_{\p}^{\ua}(z) { \widetilde \bP}_{\m}^{\ua}(z) = 2\cosh( \phi(z) + (\hB + \hmu )/2 ) \; \cosh(\phi(z) - (\hB - \hmu )/2 ).
\eea
 We can now compute the densities using (\ref{eq:exprho0}). For the horizontal wing the result is
\bea\label{eq:exprho}
e^{ 2 \hB \rho_{\ra}(z) } &=& \frac{ \tP_{\p}^{\ra}(z) \bP_{\m}^{\ra}(z)}{ \tP_{\m}^{\ra}(z) \bP_{\p}^{\ra}(z)} = \frac{\cosh(\phi(z) + (\hB + \hmu )/2 ) \; \cosh(\phi(z) + (\hB - \hmu )/2 ) }{ \cosh(\phi(z) - (\hB + \hmu )/2 )  \; \cosh(\phi(z) - (\hB - \hmu )/2 ) },
\eea
and, denoting the rhs of this equality as $e^{ 2 \hB \rho_{\ra}(z) } = R(\hB, \hmu ; z)$, the density characterising the vertical wing is given by
\bea\label{eq:exprho2}
 e^{-2\hmu \rho_{\ua}(z)} &=&  R(\hmu, \hB ; z).
\eea
The kernel appearing in the free energy formula (\ref{eq:ET4}) reduces to
\bea
\log \left(\frac{ \sinh^2( 2 \phi(z) ) \sinh^2( \hB)}{ \sinh^2( \hB \rho_{\ra}(z) )}\right) &=&\log \left(\frac{ \sinh^2( 2 \phi(z) ) \sinh^2( \hmu)}{ \sinh^2( \hmu \rho_{\ua}(z) )} \right)\nn\\
&=&-\sum_{\sigma_1, \sigma_2} \log \left(\cosh( \phi(z) +\sigma_1 \hB/2 + \sigma_2 \hmu/2 )\right),
\eea
with $\sigma_1, \sigma_2 =\pm 1$, giving the well-known result for the Gibbs free energy at $\bu=0$.

\begin{figure}
\begin{minipage}[b]{0.5\linewidth}
\centering
\includegraphics[width=7.3cm]{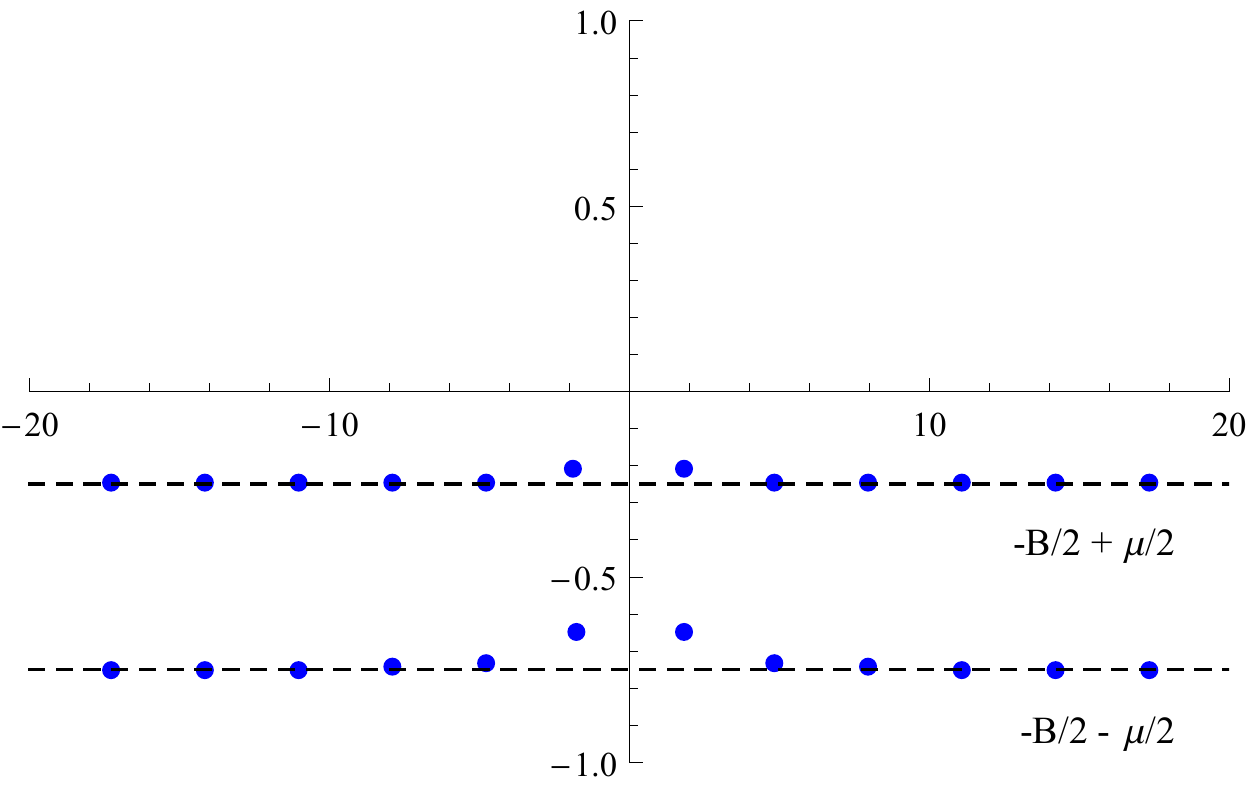}
\caption{ Position of the zeros of ${ \widetilde \bP }_{\p}^{\ra}$ in the complex $z$ plane, for the free fermion solution at  $B=1$ and $\mu=0.5$ and $T=1$.}
\label{gM}
\end{minipage}
\hspace{0.1cm}
\begin{minipage}[b]{0.5\linewidth}
\centering
\includegraphics[width=7.3cm]{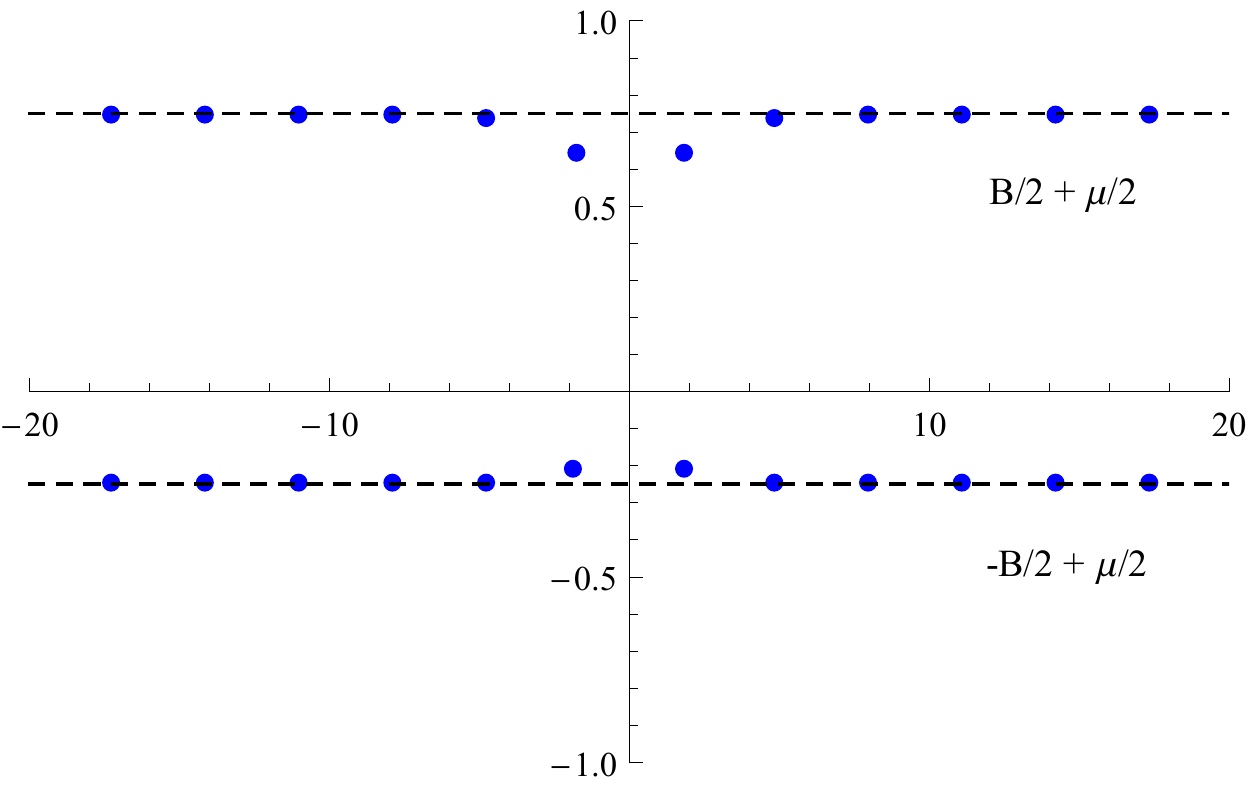}
\caption{Position of the zeros of ${ \widetilde \bP }_{\p}^{\ua}$ in the complex $z$ plane, for the free fermion solution at  $B=1$ and $\mu=0.5$ and $T=1$.}
\label{gD}
\end{minipage}
\end{figure}
The pattern of Bethe roots displayed by the free fermion solution is interesting, as the numerical solution for $\bu>0$ (see Section \ref{sec:numerics}) suggests that the zeros are smoothly deformed away from their positions at  $\bu=0$. 
Each of the $\bP$ functions has two infinite strings of zeros on the second sheet, corresponding to the two factors on the rhs of (\ref{eq:explP}). 
Denoting as $z_n( \hB, \hmu)$ the solution of the equation
\bea
2 \, \phi( z_n( \hB, \hmu) )+ \hB + \hmu = i (2 n + 1) \pi,
\eea 
and defining 
\bea
A( \hB, \hmu) = \left\{ z_n(\hB, \hmu), n \in \mathbb{Z} \right\},
\eea
the distribution of zeros is summarised in Table \ref{tab:Pzeros}, and illustrated in Figures \ref{gM} and \ref{gD} 
for two of the ${\widetilde \bP }$'s 
at $B=1$ and $\mu=1/2$. Zeros belonging to $A( \hB, \hmu)$ accumulate at infinity along the line $\IIm(z) = -B/2 - \mu/2$, and their asymptotic spacing is $\pi T$. As can be obtained from equations (\ref{eq:wronskiansfirst})-(\ref{eq:wronskianslast}) in the $\bu \rightarrow 0$ limit, in the free fermion case the zeros of the $\bF$ functions are a subset of the zeros of the $\bP$'s (see Table \ref{tab:Pzeros}).

Let us make a short comment on the analytic continuation mechanism governing the transition to excited states. As already anticipated in Section \ref{sec:energyBethe}, the free energy acquires an extra residue whenever one of the energy-carrying Bethe roots crosses the integration contour. The movement of these zeros can be driven by analytic continuation in $B$ and $\mu$. 
In general, such a crossing does not correspond to a branch point in the domain of the parameters $\hB$ or $\hmu$, since the contour can be deformed to avoid the contact with the wandering zero. Genuine branch points correspond to the so-called pinching phenomenon, when a pair of zeros collide on the contour from opposite sides~\cite{DT}. 
In the free fermion case, this happens at values of $B$ and $\mu$ given by:
\bea\label{eq:bpoints}
\lambda_B/\lambda_\mu \, e^{ \pm 2/T} = -1, \;\;\;\; \lambda_B \, \lambda_\mu \, e^{ \pm 2/T} = -1,
\eea
where $\lambda_B = e^{ B/T}$, $\lambda_{\mu}=e^{\mu/T}$ are the fugacities, corresponding to a  
pair of zeros of $\sinh^2( \hB )/\sinh^2( \hB \rho_{\ra} ) = \sinh^2(\hmu)/\sinh^2( \hmu \rho_{\ua} )$ pinching the contour of integration at the origin in the $z$-plane.
Analytic continuation 
around one of these points causes the transition to an excited branch of the free energy, and of the $\bP$ functions. 
\begin{table}[tb]
\begin{center}
\begin{tabular}{|c|c|}
\hline
Function & Zeros \\
\hline
$\bP_{a}^{\ra}(z)$  & $z \in A( \hB_a, \hmu ) \cup A( \hB_a, -\hmu)$ \\
\hline
$\bP_{a}^{\ua}(z)$  &  $z \in A( \hB, \hmu_a ) \cup A( -\hB, \hmu_a )$ \\
\hline
$\bF_{ a b }(z)$  &  $z \in A( \hB_a, -\hmu_b )$ \\
\hline
\end{tabular}
\caption{\small Distribution of the Bethe roots among different $\bP$ and $\bF$ functions, with $a, b = \p, \m$, and $\hB_{\bpm} = \pm B$, $\hmu_{\bpm} = \pm \mu$. Zeros of $\bP$'s live on the second sheet. }
\label{tab:Pzeros}
\end{center}
\end{table}
Finally, it is interesting to notice that the branch points (\ref{eq:bpoints}) mark the boundaries of the four phases of the system at $T=0$. 
In the interacting regime $\bu>0$ at $T=0$, the phase diagram includes a fifth phase describing the Mott insulator behaviour \cite{LiebWu} (see also Chapter 6 of \cite{HubbardBook}).
It would be very interesting to investigate the branching structure of the free energy at finite temperatures and coupling, and link it to the phase diagram of the Hubbard model. We expect this to be possible with the numerical method described in Section \ref{sec:numerics} and 
plan to come back to these questions in the future. 
\section{Numerical solution}
\label{sec:numerics}
In Section \ref{sec:resolvent}, we have parametrised the $\bP$ functions appearing in the problem  in terms of the densities $\rho_{\ra}$ and $\rho_{\ua}$. They can be computed by solving a system of coupled nonlinear integral equations, which determine simultaneously the densities and the function $Y_{1,1}$ entering the TBA equations. 
This formulation is very similar to the set of NLIEs proposed in \cite{FiNLIE} for the AdS$_5$/CFT$_4$ spectral problem. 
We discuss here only the ground state equations.
\subsection{Nonlinear integral equations} 
First, by expressing $Y_{1,1}$ in  the $\mT_{\ra}$ and in the $\mT_{\ua}$ gauge, we find
\bea\label{eq:Nlie2}
 r(z) &=& \frac{ \mT^{\ra\,[+]}_{1,1}(z) \; { \widetilde{ \mT^{\ra\,[-] }_{1,1} }(z) } }{\mT^{\ra\,[-]}_{1,1}(z) \; { \widetilde{ \mT^{\ra \,[+] }_{1,1} }(z) } } = e^{ 4\phi(z) } \frac{ \mT^{\ua\,[+]}_{1,1}(z) \; { \widetilde{ \mT^{\ua\,[-] }_{1,1} }(z) } }{\mT^{\ua\,[-]}_{1,1}(z) \; { \widetilde{ \mT^{\ua\,[+] }_{1,1} }(z) } },
\eea
 where 
\bea
r(z)= \left(\frac{ 1 + \hY_{1, 1}(z) \; e^{4 \phi(z)}  }{ 1 + \hY_{1, 1}(z) }\right),
\eea
and the functions $\mT_{1,1}^{i}$, $i=\bigra, \bigua$ depend on the densities through the parametrisation (\ref{eq:defmT}). In an iterative scheme, equations  (\ref{eq:Nlie2}) can be used to update the values of  the two densities starting from the knowledge of $r(z)$. The numerical method is described in \cite{FiNLIE} and is  reviewed below in Section \ref{sec:solvedensities}. To close the system, there is a further equation determining $Y_{1,1}$ as a function of $\rho_{\ra}$ and $\rho_{\ua}$:
\bea\label{eq:pvYtext}
\log{ \hY_{1,1}(z) }
&=& 2 \bu/T -2 \phi(z) + \log \left(\frac{ { \mT }_{1,2}^{ \ra }(z)}{{ \mT }_{2,1}^{ \ua }(z)} \right)+ \int_{-1}^{1} \frac{dv}{ 2 \pi i ( v - z ) }\; \disc\left[ \log \left(\frac{ \mT^{\ra\,[+1]}_{1,1}(v) \;  \mT^{\ra\,[-1]}_{1,1} (v) }{ \mT^{\ua\,[+1]}_{1,1} (v) \;  \mT^{\ua\,[-1]}_{1,1}(v) } \right)\right] \nn\\
&-& \log \left(\frac{ \mT^{\ra \,[+1]}_{1,1}(z) \; \mT^{\ra\,[-1]}_{1,1}(z) }{ \mT^{\ua\,[+1]}_{1,1}(z) \;\mT^{\ua\,[-1]}_{1,1} (z) } \right).
\eea
This relation is derived in Appendix \ref{app:proof}, both from the TBA equations and also purely from the Riemann-Hilbert formulation (\ref{eq:constraint31})-(\ref{eq:wT}). 
 An equivalent form of (\ref{eq:pvYtext}), which is convenient for the numerical implementation, is 
\bea\label{eq:pvY0}
\log{ \hY_{1,1}(z) }&=& 2 \bu/T -2 \phi(z) + \log\left(\frac{ { \mT }_{1,2}^{ \ra }(z)}{{ \mT }_{2,1}^{ \ua }(z)}\right) + \dashint_{-1}^{1} \frac{dv}{ 2 \pi i ( v - z ) }\; \disc\left[ \log\left(\frac{ \mT^{\ra\,[+]}_{1,1}(v) \;  \mT^{\ra\,[-]}_{1,1} (v) }{ \mT^{\ua\,[+]}_{1,1} (v) \;  \mT^{\ua\,[-]}_{1,1}(v) }\right) \right]\nn\\
&-& \frac{1}{2} \log \left(\frac{ \mT^{\ra \,[+1]}_{1,1}(z) \; \mT^{\ra\,[-1]}_{1,1}(z) \; { \widetilde{ \mT^{\ra \,[+1]}_{1,1} }}(z)  \; { \widetilde{ \mT^{\ra \,[-1]}_{1,1} }}(z) }{ \mT^{\ua\,[+1]}_{1,1}(z) \;\mT^{\ua\,[-1]}_{1,1} (z) \; { \widetilde{ \mT^{\ua \,[+1]}_{1,1} }}(z) \; { \widetilde{ \mT^{\ua \,[-1]}_{1,1} }}(z) } \right),
\eea
where $\dashint $ denotes Cauchy's principal value integral. 
\subsection{The numerical method}\label{sec:solvedensities}
The system (\ref{eq:Nlie2}),(\ref{eq:pvY0}) can be solved iteratively for the values of $\rho_{\ra}(z)$, $\rho_{\ua}(z)$ and $Y_{1,1}(z)$ on the interval $z \in (-1, 1)$. 
One iteration step, updating the values of the densities $\rho_{i}^{(k)} \rightarrow \rho_i^{(k+1)}$, can be represented as
\bea\label{eq:scheme}
\rho_{\ra}^{(k)}, \, \rho_{\ua}^{(k)} \longrightarrow^{(\ref{eq:pvY0})} Y_{1,1}^{(k)} \longrightarrow^{(\ref{eq:Nlie2})} {\rho_{\ra}^{(k+1)} }', {\rho_{\ua}^{(k+1)} }'.
\eea
After each step, we update the solution as $\rho_i^{(k+1)} =  \theta \rho_i^{(k)} +  (1-\theta) {\rho_i^{(k+1)} }' $. The introduction of the weights $\theta$, $1-\theta$ is a common recipe used to ensure convergence. In all cases we considered, the scheme was stable taking $\theta = \frac{1}{2}$. \\
In the above described procedure, the most difficult step is the solution of (\ref{eq:Nlie2}) for the densities. 
Let us review the method discussed in \cite{FiNLIE}, concentrating on the equation for the horizontal wing. 
Using (\ref{eq:defmT}) and the basic property $\widetilde{ \CG }_{\ra} = \CG_{\ra} + \rho_{\ra}$, the first equality in (\ref{eq:Nlie2}) can be written as 
\bea\label{eq:r}
r(z) &=&  \frac{ \sinh( \hB( \CG_{\ra}^{[2]}(z) - \slG_{\ra}(z) +\rho_{\ra}(z)/2 + 1 )) \; \sinh(\hB( { \slG_{\ra}(z)} - \CG_{\ra}^{[-2]}(z) + \rho_{\ra}(z)/2 + 1 ))  }{\sinh(\hB( \CG_{\ra}^{[+2]}(z) - \slG_{\ra}(z) -  \rho_{\ra}(z)/2 + 1 )) \; \sinh(\hB( { \slG_{\ra}(z)} - \CG_{\ra}^{[-2]}(z) -\rho_{\ra}(z)/2 + 1 )) },\nn\\
\eea
where $\slG_{\ra}$ denotes the Cauchy principal value integral
\bea\label{eq:slG0}
\slG_{\ra}(z) = \frac{1}{2 \pi i } \dashint_{-1}^{1} \frac{\brho_{\ra}(v)}{z - v } dv.
\eea
By extracting\footnote{ Notice that 
 (\ref{eq:r}) can be written as a quadratic equation for $\tanh( \hB \rho_{\ra}/2 )$.} $\rho_{\ra}$ from the rhs of (\ref{eq:r}), 
 we obtain the density in terms of $\CG^{[\pm2]}_{\ra}$, $\slG_{\ra}$ and $r$. 
 This equality is used to update the value of $\rho_{\ra}$ in the last passage of (\ref{eq:scheme}).

 The numerical evaluation of the singular integrals appearing in the NLIEs (\ref{eq:Nlie2}),(\ref{eq:pvY0}) can be performed very efficiently using a Chebyshev expansion. 
To optimize the numerical method we found it convenient to discretise the densities by using a Chebyshev expansion of the second kind
\bea\label{eq:chebyexpand}
\rho_{\alpha}^{\text{num}}(z) = \sqrt{1-z^2} \sum_{n=0}^{N_{\text{trunc}}} c^{(\alpha)}_{2n} \, U_{2n}(z), \;\;\;\; z \in (-1, 1), \;\;\; \alpha = \bigra, \bigua,
\eea
 (where we have taken the correct parity into account) and evaluate principal value integrals using the properties
\bea
\dashint_{-1}^{1} \frac{dv}{ \pi ( v - z ) }\; \sqrt{1-v^2} \, U_n(v) = -T_{n+1}(z), \;\;\;\;\dashint_{-1}^{1} \frac{dv}{ \pi( v - z ) \, \sqrt{1-v^2} } \, T_n(v) = U_{n-1}(z),\nn
\eea
where $T_n$ and $U_n$ denote the Chebyshev polynomials of the first and second kind, respectively. To produce the data presented in this paper, we took $N_{\text{trunc}} = 50$. 
In the vast majority of the cases we considered, less than 30 iterations were sufficient to achieve convergence of the coefficients $c_{n}^{(\alpha)}$ entering (\ref{eq:chebyexpand}) on the fifth digit. 
We observe that the error is approximately halved at every iteration. 

Let us also make a comment on the region of convergence. We observe that, for fixed values of $\hB$, $\hmu$ and $\bu$, the iterative scheme is convergent for sufficiently high temperatures -- in particular a preliminary study suggests that, for arbitrary $B,\mu \in \mathbb{R}$, the convergence region probably includes $0<\bu<2$ and $T \geq 1$  -- but breaks down below a certain threshold temperature. Lowering $\bu$, the breakdown temperature decreases, and this hints that, for a given value of $T$, the method should be applicable without modifications in a nonvanishing neighbourhood of the free fermion point. 
We suspect that the breakdown of the method for low temperature or strong coupling is related to the appearance of zeros on the first sheet for the ground state solution~\cite{DDVHubbard}. We plan to come back to this issue in the near future.

\begin{figure}[t]
\centering
\includegraphics[width=11cm]{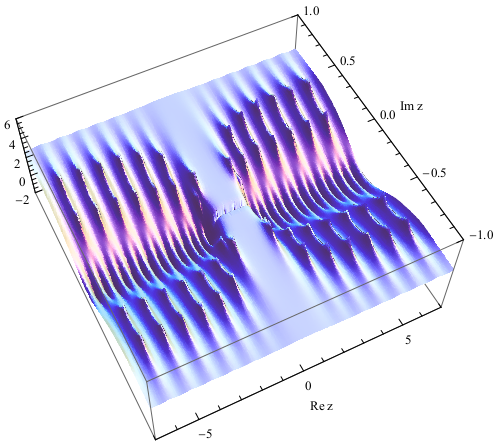}
\caption{
Plot of the function $-\RRe\left( \log( \sinh( \hB \rho_{\ra}(z) ) \right)$ in the complex $z$-plane, for $B=1$, $\mu=0.5$, $T=0.3$ and $\bu = 0.1$. 
The positive peaks correspond to energy-carrying Bethe roots, while the negative peaks on the real axis are zeros of $\sinh( 2 \phi(z) )$ (cf. equation (\ref{eq:sinhrho})).
\label{fig:contours1} 
}
\end{figure}

\subsection{Exploring the complex plane}
The numerical method we have described computes $\rho_{\ra}(z)$, $\rho_{\ua}(z)$ and $Y_{1,1}(z)$ for $-1 < z <1$. 
Once we have a solution on the interval, we can reconstruct the behaviour of these functions in the complex plane. 
We do this in two steps. First, we use equation (\ref{eq:pvYtext})
to compute $Y_{1,1}(z)$ for $z$ on the whole first Riemann section -- with cuts  at $z \in (-1,1 )$, $z \in (-1,1) \pm 2 i \bu$ -- from the values of the densities on the interval.
From the same information, we can also compute  $\CG_{\ra}(z)$ and $\CG_{\ua}(z)$ for an arbitrary complex value of $z$. 
Then, one can obtain the values of $\rho_{\ra}(z)$, $\rho_{\ua}(z)$ for any complex values of $z$ by inverting (\ref{eq:Nlie2}). The complex zeros of $\sinh(\rho_{\ra}(z) )$ are visible in Figure \ref{fig:contours1} for the numerical solution corresponding to $\mu=1/2$, $B=1$, $\bu=0.1$, showing a clear qualitative similarity with the free fermion case. 

\section{Mirror equations}
\label{sec:mirror}

In this Section we shall describe the finite-size versions of the functional relations (\ref{eq:closed2})-(\ref{eq:closed4}), 
 which form a set of equations fully equivalent to the Lieb-Wu quantisation conditions (\ref{eq:Lieb-Wu}),(\ref{eq:Lieb-Wu1})  with more general twisted boundary conditions. 
 Comparing the thermal BA (\ref{eq:BA00}),(\ref{eq:BA0}) and the Lieb-Wu  equations (\ref{eq:Lieb-Wu}),(\ref{eq:Lieb-Wu1}), and considering the dispersion relations implied by  (\ref{eq:Betheenergy}), (\ref{eq:ET4}), one can see that the two systems are formally related by a simple map swapping $L \leftrightarrow 1/T$ and single-particle energies with momenta\footnote{ 
 Notice that the identification of $i \thh(z) $ as the free energy carried by a Bethe root of rapidity $z$ is motivated by the discussion of Section \ref{sec:energyBethe}. 
}:
 \bea\label{eq:mirrormap}
\begin{array}{ccc} ( \thh(z) , \beps(z) ) & \longrightarrow & ( -i \beps(z)  , i \thh(z) ) \\
 & \text{ $L \rightarrow 1/T $} & .
\end{array}
 \eea 
  Applying (\ref{eq:mirrormap}) to (\ref{eq:closed2})-(\ref{eq:closed4}), 
 we find a simple set of functional relations:
\bea
\bq_{\p}^{\ra}(z) { \widetilde \bq}^{\ra}_{\m}(z) - { \widetilde \bq}^{\ra}_{\p}(z) \bq^{\ra}_{\m}(z) &=&   2 i \;\sin( L \, \thh(z) ),\label{eq:closed2mir} \\
\bq^{\ua}_{\p}(z) { \widetilde \bq}^{\ua}_{\m}(z) - { \widetilde \bq}^{\ua}_{\p}(z) \bq^{\ua}_{\m}(z) &=& -2 i \; \sin( L \, \thh(z) ),\label{eq:closed3mir} \\
{ \widetilde \bq }_a^{\ra}(z) \bq^{\ua}_b(z) = \bG_{ab}^{[+1]}(z) e^{ -\frac{i}{2} \thh(z) \, L } &+& \bG_{ab}^{[-1]}(z) e^{ \frac{i}{2} \thh(z) \, L}, \;\;\;a, b = \p, \m,\label{eq:closed4mir}
\eea
where we require that the functions $\bq^{\alpha}_{a}(z)$ live on a two-sheeted Riemann surface, while the $\bG_{a b}$'s  are entire. 
 Relations (\ref{eq:closed2mir})-(\ref{eq:closed4mir}) can be seen as a set of Baxter-like equations for 
 the Hubbard Hamiltonian. 
 For $L$ an even integer, they admit many solutions where the $\bG$'s are polynomials; consequently, the $\bq$'s can be written as polynomial functions of $\x(z)$ and ${\widetilde \x}(z)$, fixed in terms of their zeros on two sheets. For completeness, we can also consider exponential prefactors of the same type as the ones in (\ref{eq:Q1}), relabeling $B \rightarrow \alpha$, $\mu\rightarrow \beta$.
 We then find easily the quantisation conditions for the following Bethe parameters: the zeros $s_j$ of the $\bq$'s, and $\lambda_{l}$ of the $\bG$'s.  Setting $\thh(s_j) = k_j$, and following the same route of Section \ref{sec:QTM}, we find
\begin{align} 
e^{i k_j L + i (\halpha - \hbeta) } &= -\prod_{l=1}^{M} \left(\frac{\lambda_l-\sin(k_j) - i\bu}{\lambda_l-\sin(k_j) + i\bu}\right),\;\;  j \in \left\{1,\dots,N\right\},  \label{eq:twistedBA0}\\  
e^{-2 i\hbeta  } \, \prod_{j=1}^{N} \left(\frac{\lambda_l-\sin(k_j) - i\bu}{\lambda_l-\sin(k_j) + i\bu}\right) &=
\prod_{^{m=1}_{m\ne l} }^{M} \left(\frac{\lambda_l-\lambda_m - 2i\bu}{\lambda_l-\lambda_m + 2i\bu} \right),\;\;  l \in \left\{1,\dots,M\right\}, \label{eq:twistedBA1}
\end{align} 
which is precisely the BA diagonalising the Hubbard Hamiltonian on the $L$-site chain  with twisted boundary conditions~\cite{HubbardTwisted, HubbardToroidal}:
\bea
c_{L+1, \uparrow} = c_{1, \uparrow} \, e^{i (\halpha - \hbeta - \pi)}, \;\;\;\; c_{L+1, \downarrow} = c_{1, \downarrow} \, e^{i (\halpha + \hbeta - \pi ) }.
\eea
 The relation between the two Bethe Ansatz systems discussed above is directly connected to the path integral approach to the thermodynamics. 
 Adopting the notation of \cite{QTMReview}, the latter is based  on rewriting the partition function,
 \bea\label{eq:defiZ}
 Z_{\text{1D quantum}} = \text{tr}_{ V_{\text{phys}} } \, e^{-1/T \, \mathcal{H} } ,
\eea
of a spin chain of $L$ sites with Hamiltonian $\mathcal{H}$, as
\bea
 Z_{\text{1D quantum}} = \lim_{\NT\rightarrow \infty} \left. Z_{\text{2D classical} }( \NT, L , u  ) \right|_{ u= \frac{1}{T \NT}},
 \eea
 where $Z_{\text{2D classical} }( \NT, L , u )$ is the partition function of an appropriately defined 
 two-dimensional statistical model living on a $L \times \NT$ lattice.
 Its column-to-column transfer matrix is the Quantum Transfer Matrix $T_{\text{QTM}}(u)$. 
  By a $\pi/2$ rotation on the lattice we switch from (\ref{eq:defiZ}) to the description
 \bea
 Z_{\text{1D quantum}} = \lim_{\NT \rightarrow \infty} \text{tr}_{V_{\text{Trotter} } } \left. (T_{\text{QTM}}(u) )^L \right|_{ u=\frac{1}{T \NT} } .
 \eea
 The exact Bethe Ansatz (\ref{eq:BA00}),(\ref{eq:BA0}) describes the spectrum of $T_{\text{QTM}}$ in the Trotter limit 
 $\NT \rightarrow \infty$. It is an interesting open problem whether one can define the QTM directly in this limit, and give it a meaningful physical interpretation as a continuum model living on a space of size $1/T$.

A somehow similar problem has arisen in the context of AdS/CFT integrability, where Zamolodchikov's ideas on the TBA for Lorentz invariant scattering  theories \cite{ZamolodchikovTBA} were adapted to the study of the non-relativistic string sigma model dual to planar $\mathcal{N}{=}4$ SYM by considering its doubly Wick rotated counterpart (the  \emph{mirror model}) \cite{Wrapping, ArutyunovMirror}. 
 Thanks to the knowledge of the action for the $AdS_5 \times S^5$  string sigma model, it has recently been possible to identify the corresponding mirror model as a string theory living on a mirror background \cite{MirrorString}.
 
 Finally, we would like to mention that is also possible to introduce a mirror version of Takahashi's TBA, in such a way that it is equivalent to the finite-size BA (\ref{eq:twistedBA0}),(\ref{eq:twistedBA1}). 
 Apart from minor subtleties \footnote{For example, we think the $-2\bu/T$ term in (\ref{eq:tba3})
  should be dropped.}, the mirror TBA equations can be obtained from (\ref{eq:tba1})-(\ref{eq:tba3}) and the energy formula (\ref{eq:energyrw}) through the following formal map :
\bea
\beps(z) &\rightarrow& i \thh(z),  \;\;\;\; \thh(z) \rightarrow -i \beps(z) , \;\;\; T \rightarrow 1/L, \;\;\;\; \{ \hB , \hmu \}\rightarrow \{ i\halpha, i\hbeta\}, 
\eea
and by modifying the integration contours as (cf \cite{ArutyunovMirror, BFT, GKKV, AF})
\bea
\int_{-1 + i 0^+}^{1+i 0^+} \rightarrow \int_{-\infty + i 0^+}^{-1 + i 0^+} +  \int_{1+i 0^+}^{\infty + i 0^+} ,
\eea 
 so that the roles of the mirror and magic sheets of Figures \ref{fig:magicsheet}, \ref{fig:mirrorsheet} are interchanged\footnote{Due to this change in kinematics, the mirror TBA described here has some features in 
 common with the TBA for the B model of \cite{FrolovQuinn}, although the latter has a different dispersion relation.
 It would be interesting to investigate the relation between the two. }. From these equations, one could in principle repeat the reduction presented in this paper and obtain the system (\ref{eq:closed2mir})-(\ref{eq:closed4mir}).  

\section{Conclusions}
\label{sec:conclusions}
The one-dimensional fermionic Hubbard model is one of the most interesting systems of low-dimensional condensed matter physics. Since its appearance in 1963, it has been intensively studied by means of exact and perturbative methods, greatly advancing the  understanding of the physics of electron transport in 1D solids.
A partial grasp about the huge number of results on this model can be obtained by consulting \cite{DeguchiReview}, the book \cite{HubbardBook} and the collections of works in \cite{MontorsiHubbard,HubbardBook2}. The purpose of this article is to add a little piece to the jigsaw, by recasting the Thermodynamic Bethe Ansatz equations of Takahashi as a nonlinear Riemann-Hilbert problem, reminiscent of the Quantum Spectral Curve formulation recently obtained for the study of anomalous dimensions in AdS/CFT~\cite{QSC}. 
One of the main results presented in this paper is a new set of nonlinear integral equations describing the thermodynamics of the system. 
 In their region of validity (see discussion at the end of Section \ref{sec:solvedensities}), these equations can be integrated numerically with very high precision and, even when implemented on Mathematica, the iterative algorithm converges in only a few seconds of CPU time. 
The complexity of this formulation is comparable to the system of nonlinear integral equations derived by J\"uttner, Kl\"umper and Suzuki in \cite{DDVHubbard}. However, as a consequence of the fact that the equations proposed here are defined on a finite support, we think that they may prove more convenient for the study of finite temperature correlation lengths. 
 
\medskip
\noindent{\bf Acknowledgments --}
We are especially indebted to Nikolay Gromov and Junji Suzuki for many important suggestions and help. 
We also thank Davide Fioravanti, Fedor Levkovich-Maslyuk and Stefano Negro for many discussions and past 
collaboration on related topics, Stijn van Tongeren for useful correspondence and Volker Schomerus for encouragement and interest in the project. 
The research leading to these results has received funding from the People Programme (Marie Curie Actions) of the European Union's Seventh Framework Programme FP7/2007-2013/ under REA Grant Agreement No 317089, from the INFN grant FTECP and the UniTo-SanPaolo research  grant Nr TO-Call3-2012-0088 {\it ``Modern Applications of String Theory'' (MAST)}.

\appendix
\section{The magic sheet}
\label{sec:magic}
In this Section we will show that the Y functions have the following cut structure on the magic sheet: 
\begin{itemize}
\item  $Y_{1, n}(z)$ and $Y_{n, 1}(z)$  for $n \geq 2$ have only four branch cuts: 
$$ 
z \in(-1, 1) \pm i \bu n, \;\;\;\; z \in (-1, 1) \pm i \bu (n+2),
$$ 
\item $Y_{1,1}(z) $ and $\hY_{2,2}(z)$ have only three branch cuts: 
$$ 
z \in (-1, 1), \;\;\;\; z \in (-1, 1) \pm 2 i \bu.
$$
\end{itemize}
As we will argue in Appendix \ref{app:twosheets}, a stronger property is true, namely the Y functions have no branch points outside the positions specified above, on any Riemann sheet. However, here we will present only the proof on the magic sheet, which is easier as it relies only on the structure of the Y-system and discontinuity relations. Before we start, we need to rewrite the Y-system (\ref{eq:Ysys}) -- originally defined on the mirror section -- on the magic sheet. Using (\ref{eq:disco1})-(\ref{eq:dis4}), it is simple to prove that the magic-sheet version of the Y-system is
\bea\label{eq:modY}
Y_{1,n}^{[+1]} \; Y_{1,n}^{[-1]} &=& ( 1 + Y_{1, n+1} )( 1 + Y_{1, n-1} ),  n \geq 2\\
 \frac{1}{Y_{n, 1}^{[+1]} \; Y_{n, 1}^{[-1]}} &=& ( 1 + 1/Y_{ n+1, 1} )( 1 + 1/Y_{n-1, 1} ), \;\;\; n \geq 2, \\
 \frac{ Y_{1,1}^{[+1]} }{ Y_{2,2}^{[-1]} } &=& \left(\frac{ 1 + Y_{1, 2} }{ 1 + 1/Y_{2, 1}  } \right).
\eea
 We will now test the cut structure of the $Y_{1, n+1}$ functions for $n \geq 1$.  First, let us introduce
\bea
X_n = \left(\frac{ 1 + \hY_{1, n}^{[n+1]} }{ \hY_{1, n}^{[n+1]} }\right) 
\left( \frac{ 1 + \hY_{1, n+1}^{[n]} }{ \hY_{1, n+1}^{[n]} }\right), \;\;\;\;\; n \geq 2.
\eea
These combinations are useful since, applying (\ref{eq:modY}) in two elementary steps, it is possible to prove that
\bea\label{eq:recur}
\frac{\hY_{1, n}^{[n+3]} }{ (1 + \hY_{1,n-1}^{[n+2]})^{1-\delta_{n, 2} } } = X_{n} \; X_{n+1} \left( \frac{ 1 + \hY_{1,n+3}^{[n]}  }{ \hY_{1,n+2}^{[n-1]}}\right), \;\;\;\; n \geq 2.
\eea
Notice that the term in brackets in (\ref{eq:recur}) has no cut because it falls into the analyticity strip. Therefore,  
the cut structure of the lhs of (\ref{eq:recur}) depends on the $X_n$ factors. We shall now show that $X_n$ have no cut on the real axis, for all $n \geq 2$. 
Starting from $n = 2$, we can compute
\bea
X_2 = \frac{(1 + \hY_{1, 2}^{[3]} ) (1 + \hY_{1, 3}^{[2]} ) }{ \hY_{1, 2}^{[3]} \hY_{1, 3}^{[2]} } = \frac{ \hY_{1, 2}^{[1]} (1 + \hY_{1, 2}^{[3]} )   }{ \hY_{1, 3}^{[2]}  } = \frac{ ( \hY_{1, 2}^{[1]} + 1 +  \hY_{1, 3}^{[2]} )   }{ \hY_{1, 3}^{[2]}  } = 1 + \frac{ \hY_{1, 3} }{ (1 + \hY_{1, 4}^{[1]} ) }.
\eea
This expression manifestly does not have branch points on the real axis, as all terms on the rhs fall into their respective analyticity strips.
A very similar calculation shows that
\bea
X_3 &=& \frac{(1 + \hY_{1, 3}^{[4]} ) (1 + \hY_{1, 4}^{[3]} ) }{ \hY_{1, 3}^{[4]} \hY_{1, 4}^{[3]} } = \frac{ \hY_{1, 3}^{[2]} (1 + \hY_{1, 3}^{[4]} )   }{ \hY_{1, 4}^{[3]} (1 + \hY_{1, 2}^{[3]} )  } = \frac{  \hY_{1, 3}^{[2]} + ( 1 +  \hY_{1, 4}^{[3]} )( 1 +  \hY_{1, 2}^{[3]} ) }{ \hY_{1, 4}^{[3]} (1 + \hY_{1, 2}^{[3]} )  } \nn\\
&=& 1 + \frac{ \hY_{1, 4}^{[1]} ( \hY_{1, 3}^{[2]} +  1 +  \hY_{1, 2}^{[3]} ) }{ (1 + \hY_{1, 5}^{[2]} )  (1 + \hY_{1, 3}^{[2]} ) (1 + \hY_{1, 2}^{[3]} ) } =  1 + \frac{ \hY_{1, 4}^{[1]} }{ (1 + \hY_{1, 5}^{[2]} ) } 
- \frac{1}{ X_2 } \; \frac{ \hY_{1, 4}^{[1]} }{ ( 1 + \hY_{1, 5}^{[2]} ) },
\eea
and, literally by repeating this calculation with shifted indices, one finds the general case:
\bea
X_{n+1} =  1 + \left(\frac{ \hY_{1, n+1}^{[n-2]} }{ 1 + \hY_{1, n+2}^{[n-1]}  } \right) - \frac{1}{ X_{n} } \left(\frac{ \hY_{1, n+2}^{[n-1]} }{  1 + \hY_{1, n+3}^{[n]}  }\right), \;\;\;\;\;n\geq 2.
\eea
Taking into account the analyticity strips of the functions on the rhs, this equation shows by induction that all $X_n$ are free of cuts. 
 Therefore, the lhs of (\ref{eq:recur}) is analytic in a neighbourhood of the real axis, meaning that  $Y_{1,n}$ with $n \geq 2$ has possibly two cuts in the upper half plane at $\IIm(z) = \bu (n \pm 1)$, but no cut at $\IIm(z) = \bu (n+3)$. Moreover, because of the Y-system, no further cuts are possible and    the $Y_{1,n}$'s with $n \geq 2$ have only two short branch  cuts in the upper half plane.
By symmetry, the argument can be repeated for the lower half plane and for the $\hY_{n, 1}$ functions with $n \geq 2$. \\
 The argument presented above (contrary to the remaining part of this Appendix) can be straightforwardly adapted to the horizontal wing of the AdS/CFT Y-systems.

Let us now prove that $\hY_{1,1}$ and $\hY_{2,2}$ have only three cuts. 
 Using (\ref{eq:modY}), we can compute
\bea
\hY_{1,1}^{[4]}&=& \hY_{2,2}^{[2]} \; \left(\frac{1 + \hY_{1,2}^{[3]}}{1 + 1/\hY_{2,1}^{[3]}} \right) 
= \hY_{2,2}^{[2]} \;\left(\frac{ 1 + (1 + \hY_{1,3}^{[2]})/\hY_{1,2}^{[1]}  }{ 1 + (1 + 1/\hY_{3,1}^{[2]}) \hY_{2,1}^{[1]}  } \right) \\
&=& \frac{ \hY_{2,2}^{[2]} }{ \hY_{1,2}^{[1]} \hY_{2,1}^{[1]} } \; \left(\frac{ 1 + \hY_{1,2}^{[1]}  }{  1 + 1/\hY_{2,1}^{[1]}  } \right) \,R = 
\frac{ \hY_{1,1}^{[2]}  \hY_{2,2}^{[2]}}{ \hY_{1,2}^{[1]} \hY_{2,1}^{[1]} \hY_{2,2}  } \; R,
\eea
where $R$ manifestly has no cut on the real axis and is defined by
\bea
R=  \left( \frac{1 + Y_{1,3}^{[2]}/(1 + \hY_{1,2}^{[1]} ) }{1 + (Y_{3,1}^{[2]} )^{-1}/(1 + 1/\hY_{2,1}^{[1]} ) }\right) =\left( \frac{1 + (1 + \hY_{1,4}^{[1]} )/\hY_{1,3}}{1 + (1 +  1/\hY_{4,1}^{[1]} )\hY_{3,1}} \right).
\eea
Because $Y_{1,1}(z) Y_{2,2}(z) = e^{- 4 \, \phi(z) }$ has no cuts outside the real axis, 
the only discontinuity can come from the factor:
\bea
H = 1/( \hY_{1,2}^{[1]} \hY_{2,1}^{[1]} \hY_{2,2}  ).
\eea
Using the discontinuity relations (\ref{eq:disco1})-(\ref{eq:dis4}), we find
\bea
\frac{ {\widetilde H} }{H} = \frac{ 1 + 1/\hY_{1,1} }{ 1 + \hY_{2,2} } \; \frac{ 1 + 1/\hY_{2,2} }{ 1 + \hY_{1,1} } \; \frac{ \hY_{2,2} }{ { \widetilde \hY }_{2,2} } = 
1.
\eea
This shows that $\hY_{1,1}$ (and therefore also $\hY_{2,2}$) does not have a branch cut with $\IIm(z) = 4 \bu$. The Y-system, together with the results already obtained for the other Y functions, imply that no further cuts are possible.
\section{Monodromy properties of the $\bP$ functions}\label{app:twosheets}

The purpose of this Appendix is to provide a proof for some statements made in Section \ref{sec:Tsystem}.  
\begin{enumerate}[1)]
\item  First, we prove that it is possible to choose the $\wT$ gauges in such a way that equation (\ref{eq:T10wr}) holds, namely
\bea\label{eq:wT1010}
\wT_{1,0}^{\ra}(z) = e^{-2 \phi(z) }, \;\;\;\;\;  \wT_{0, 1}^{\ua}(z) = e^{2 \phi(z) }.
\eea
In the proof we will use the resolvent parametrisation of Section \ref{sec:resolvent},  which, rigorously speaking, is valid only for the ground state. However, we expect that the result holds in general. 
\item Secondly, we derive the constraints (\ref{eq:constraint31}), (\ref{eq:constraint32}).
\item Finally, we discuss how to infer that 
the second-sheet evaluation of the $\bP$ functions, ${\widetilde \bP}$, do not have other branch cuts apart from $z \in (-1,1)$.
This shows that the $\bP$'s live on a Riemann surface with only two sheets.
\end{enumerate}
 To prove these properties, we will adapt many of the arguments of \cite{FiNLIE} to the present case. \\
\noindent
\subsection*{Proof of equation (\ref{eq:T10wr}) for $\wT_{1,0}^{\ra}$ and $\wT_{0,1}^{\ua}$ }
To verify the statement 1), let us start from the gauge $\mT^{\ra}$ (we restrict to the horizontal wing since the situation is clearly analogous for the vertical wing).  
This gauge is defined by the parametrisation
\bea\label{eq:mTpar}
 \mT^{\ra}_{1,n} &=& \sinh\left( \hB ( 1 + \CG^{[+n]}_{\ra} - \CG^{[-n]}_{\ra}  ) \right)/\sinh( \hB), \;\;n \geq 1, \nn\\
 { \mT }^{\ra }_{0,s} &=& 1, \;\; s \in \mathbb{N}, \;\;\;\;\;\;\;\; { \mT }^{\ra }_{2,l} = { \mT }^{\ra \, [+l]}_{1,1} \; { \mT }^{\ra \,[-l]}_{1,1}, \;\; l \geq 2. 
\eea
with the resolvent density $\rho_{\ra}$ 
 fixed uniquely, in terms of the Y functions, through equation (\ref{eq:Nlie2}). 
We shall prove that\footnote{The analogous expression for the upper wing is \bea\label{eq:T10up2}
\mT_{0,1}^{\ua}(z) = \frac{\sinh(\hmu\rho_{\ua}(z) )}{ 2\sinh(2 \sq(z)) \sinh(\hmu) } \; e^{2 \sq(z) }. \nn
\eea
}
\bea\label{eq:T1010}
\mT_{1,0}^{\ra}(z) = \frac{\sinh(\hB \rho_{\ra}(z) )}{2 \sinh( 2 \sq(z) ) \sinh( \hB)  } \; e^{-2 \sq(z) }. 
\eea
We start by rewriting some of the discontinuity relations (\ref{eq:disco1})-(\ref{eq:Ysys}) in terms of the $\mT^{\ra}$ functions.
Using the properties of this gauge, equation $Y_{1,1} = 1/\widetilde{Y}_{2,2}$ can be written as
\bea\label{eq:eq11}
{\widetilde{\mT}_{1,0}^{\ra} } = \mT_{3,2}^{\ra}/\mT_{2,3}^{\ra} \, \left(\frac{ { \widetilde \mT }_{2,1}^{\ra} }{ \mT_{2,1}^{\ra} } \right),
\eea
while the relation $Y_{1,1} \, Y_{2,2} = e^{-4 \phi } $ becomes
\bea\label{eq:eq22}
e^{-4 \phi } = \frac{ \mT_{1,0}^{\ra} \mT_{2,3}^{\ra} }{\mT_{3,2}^{\ra} } =  \frac{ \mT_{1,0}^{\ra} }{{\widetilde{\mT_{1,0}^{\ra} } } } \,  \left(\frac{ { \widetilde \mT }_{2,1}^{\ra} }{ \mT_{2,1}^{\ra} } \right).
\eea
The equation involving $Y_{1,2}$ is automatically satisfied by the parametrisation (\ref{eq:mTpar}), while the equation for $Y_{2,1}$ will be used later. Now, we compare the T-system equations at the nodes $(1,1)$ and $(2,2)$. On the magic sheet, these two equations read 
\bea\label{eq:Tcompare}
{ \widetilde{ \mT_{1,1}^{\ra \, [+1] } } } \, \mT_{1,1}^{\ra \, [-1]} &=& \mT_{1,0}^{\ra} \mT_{1,2}^{\ra} + \mT_{2,1}^{\ra},\\
 { \widetilde{\mT_{2,2}^{\ra \, [+1]} }} \, \mT_{2,2}^{\ra \, [-1]} &=&  \mT_{1,1}^{\ra \, [+1]} \, {\widetilde{ \mT_{1,1}^{\ra \, [-1]} } } \, \mT_{2,3}^{\ra} = \mT_{2,3}^{\ra} \mT_{2,1}^{\ra} + \mT_{3,2}^{\ra} \mT_{1,2}^{\ra}. \label{eq:Tcompare2}
\eea
 The identity (\ref{eq:Tcompare2}) was derived using $\mT_{2,s}^{\ra} = \mT_{1,1}^{\ra \,[+s]} \, \mT_{1,1}^{\ra \,[-s]}$. Eliminating $\mT_{21}^{\ra}$ from these equations shows that
 \bea\label{eq:discTT}
 \disc\left[ { \widetilde{ \mT_{1,1}^{\ra \, [+1] } } } \, \mT_{1,1}^{\ra \, [-1]} \right] = \mT_{1,2}^{\ra} \left( \mT_{1,0}^{\ra} - \mT_{3,2}/\mT_{2,3}^{\ra} \right) =  \mT_{1,2}^{\ra} \, \mT_{1,0}^{\ra} \left(1 - e^{4 \phi} \right),
 \eea
 where we have used (\ref{eq:eq22}) in the last step. 
Using (\ref{eq:mTpar}) in (\ref{eq:discTT}), we find the result quoted in (\ref{eq:T1010}). 
 To construct the gauge $\wT^{\ra}$, we can now introduce the gauge transformation factor $\bh_{\ra}$ through equation (\ref{eq:integralh}), so that
 $ \bh_{\ra} { \widetilde \bh }_{\ra} =  \sinh( 2 \sq  \sinh( \hB) )/\sinh(\hB \rho_{\ra} ) $.
 Defining the $\wT^{\ra}$ gauge through 
 \bea\label{eq:wTprop}
 \wT^{\ra}_{1,s} = \bh_{\ra}^{[s]} \bh_{\ra}^{[-s]} \, \mT_{1, s}^{\ra}, \;\;\; s \in \mathbb{N}^+, \;\;\;\; \wT_{0, n}^{\ra} = 1, \;\;\;n \in \mathbb{N}, 
 \eea
 we finally arrive at
 \bea\label{eq:T10final}
 \wT_{1,0}^{\ra}(z) = \bh_{\ra}(z) { \widetilde \bh }_{\ra}(z) \, \mT_{1,0}^{\ra}(z) = e^{-2 \sq(z) }.  
\eea
\subsection*{Proof of the constraints (\ref{eq:constraint31}),(\ref{eq:constraint32})}
Statement 2) can be proved by revisiting the derivation of (\ref{eq:discTT}) given above  
in the case of the gauge $\wT^{\ra}$. This leads to 
\bea\label{eq:impoapp} \text{disc}\left[{ \widetilde{ \wT_{1,1}^{\ra \, [+1] } } } \, \wT_{1,1}^{\ra \, [-1]}\right] =  \wT_{1,2}^{\ra} \, \wT_{1,0}^{\ra} \left(1 - e^{4 \phi} \right).
\eea 
 The constraint (\ref{eq:constraint31}) follows from (\ref{eq:impoapp}) using  
(\ref{eq:wT1010}) and the expression of the $\wT^{\ra}$ functions in terms of $\bP^{\ra}$'s.
One can also derive an equation analogous to (\ref{eq:impoapp}) with $\bigra \rightarrow \bigua$, $\phi \rightarrow - \phi$, and prove (\ref{eq:constraint32}) by the same method.
\subsection*{Proof that the $\bP$ functions have only a single cut on the second sheet} 
Statement 3): let us now sketch the proof that the $\bP$'s have no cut outside the real axis even on their second Riemann sheet, and, therefore, live on a two-sheeted surface. The proof is very similar to the ones used in a more complicated context in \cite{AdS5Long,ABJMnoi2}.
 To start, we observe that, due to (\ref{eq:T1010}), the ratio in brackets in (\ref{eq:eq11}) and (\ref{eq:eq22}) is actually one, so that ${ \widetilde \mT}_{2,1}^{\ra} = \mT^{\ra}_{2,1}$. Since $\wT_{2, 1}^{\ra} = \bh_{\ra}^{[+2]} \; \bh_{\ra}^{[-2]} \; \bh_{\ra} \; { \widetilde{\bh}_{\ra} } \, \mT_{2, 1}^{\ra} $, we find that also $\wT^{\ra}_{2,1}$ has no cut on the real axis: 
\bea\label{eq:T21nocut}
\wT_{2,1}^{\ra} = { \widetilde \wT }_{2, 1}^{\ra}. 
\eea
Furthermore, notice that we have not used yet the discontinuity equation (\ref{eq:dis4}) involving $Y_{2,1}$. Rewriting this relation in terms of T functions in the gauge $\wT^{\ra}$, we find
\bea\label{eq:discY21}
\frac{ { \wT_{1,1}^{\ra \, [+1] } }  \, \wT_{1,1}^{\ra \, [-1]} }{ \wT_{2,2}^{\ra \, [+1] }  \, \wT_{2,2}^{\ra \, [-1]} } = \frac{ \wT_{1,0}^{\ra} }{\wT_{3,2}^{\ra} } \, \frac{ \widetilde{\wT_{3,1}^{\ra \, [+1] } } \, {\wT_{2,0}^{\ra \,[+1] }} }{ \wT_{3,1}^{\ra \, [+1] } \, \widetilde{\wT_{2,0}^{\ra \,[+1] }} }. 
\eea
The ratio on the lhs side can be simplified using $\wT_{2, 2}^{\ra} = \wT_{1,1}^{\ra [+2]} \wT_{1,1}^{\ra [-2]}$, and leads to
\bea\label{eq:aaa}
1 = \frac{ \wT_{1,0}^{\ra} \, \wT_{2,3}^{\ra} }{\wT_{3,2}^{\ra} } \, \frac{ \widetilde{\wT_{3,1}^{\ra \, [+1] } } \, {\wT_{2,0}^{\ra \,[+1] }} }{ \wT_{3,1}^{\ra \, [+1] } \, \widetilde{\wT_{2,0}^{\ra \,[+1] }} } = e^{-4 \phi}\, \, \frac{ \widetilde{\wT_{3,1}^{\ra \, [+1] } } \, { \wT_{2,0}^{\ra \,[+1] }} }{ \wT_{3,1}^{\ra \, [+1] } \, \widetilde{\wT_{2,0}^{\ra \,[+1] }} }.
\eea
Using the T-system and (\ref{eq:wT1010}), it is simple to compute the factor involving $\wT_{2, 0}^{\ra}$, and finally (\ref{eq:aaa}) reduces to the condition
\bea\label{eq:T31nocut}
\wT_{3,1}^{\ra \, [+1] } = \widetilde{\wT_{3,1}^{\ra \, [+1] } }.
\eea
Furthermore, using the discontinuity relations
 \bea
Y_{n +1, 1}^{[n]}/\widetilde{  Y_{n+1, 1}^{[n]} }= \left(1 + 1/Y_{n, 1}^{[n-1]}  \right)/ \left( 1 + 1/\widetilde{ Y_{n, 1}^{[n-1]} }  \right),\;\;\;n \in \mathbb{N}^+,
 \eea
 which follow from the Y-system, it is possible to generalise (\ref{eq:T21nocut}) and (\ref{eq:T31nocut}) to 
\bea\label{eq:Tn1nocut}
\disc\left[  \wT_{n, 1}^{\ra \, [n-2]} \right] = 0, \;\;\; n=2, 3, \dots. 
\eea
The set of equations (\ref{eq:Tn1nocut}) can be used to prove that the $\bP$'s live on a two-sheeted Riemann surface. 
By definition, these functions have a single cut on their first sheet, and parametrise some of the $\wT^{\ra}$'s as
\bea\label{eq:rwapp}
\wT_{1, s}^{\ra} = \bP_{\p}^{\ra \, [+s]} \bP_{\m}^{\ra \, [-s]} - \bP_{\p}^{\ra \, [-s]} \bP_{\m}^{\ra \, [+s]}, \;\;\;  s \in \mathbb{N}^+.
\eea
Hirota equation (\ref{eq:Tsystem}) allows to extend (\ref{eq:rwapp}) and compute any $\wT^{\ra}$ function in terms of the $\bP^{\ra}$'s. However, since (\ref{eq:Tsystem}) is defined on the mirror section, the computation may require to explore the values of the $\bP$ functions on the second sheet. In particular, it can be easily checked that $\wT^{\ra, [n-2]}_{n, 1}$ with $n > 1$ involves (linearly) the values of ${ \widetilde \bP }^{\ra, [2 n-2]}_{\bpm}$, so that imposing the constraints (\ref{eq:Tn1nocut}) gives information on the possible cuts of the ${\widetilde \bP}^{\ra}$'s. 
For instance, from equation (\ref{eq:T21nocut}), one can recover the constraint (\ref{eq:constraint31}). With the aid of this result, the other equations in (\ref{eq:Tn1nocut}) can be used to prove that 
\bea
{ \widetilde{ \bP } }_{\bpm}^{\ra \,[+n]} 
 -   { \widetilde{  { \widetilde{ \bP } }_{\bpm}^{\ra \,[+n]}  }  }= 0, \;\; n \in \mathbb{N}^+.
\eea
Obviously, by symmetry all these results are valid both in the upper and in the lower half complex plane, and show that ${\widetilde \bP }^{\ra}_{\bpm}$ have no cuts outside the real axis.

\section{Fixing the gauge factor $\wT^{\ra}/\wT^{\ua}$}
\label{app:proof}
The main purpose of  this Appendix is to derive the form of the gauge transformation (\ref{eq:wTgen1})-(\ref{eq:wTgen3}) relating the $\wT^{\ra}$ and $\wT^{\ua}$ 
functions. Along the way, we will also obtain the two NLIEs (\ref{eq:rmatch}) and (\ref{eq:pvY}), adopted for the numerical solution method described in Section \ref{sec:numerics}. \\
Let us briefly discuss the logic of this Section. 
 In Appendix \ref{app:twosheets}, we have encoded the discontinuity equations (\ref{eq:disco1})-(\ref{eq:Ysys}) in the simple statement that the $\bP$ functions live on a two-sheeted Riemann surface and satisfy the relations (\ref{eq:constraint31}),(\ref{eq:constraint32}). 
However, some information is still missing to close the system. In fact, notice that, when describing the system with the functions $\bP^{\ra}_{\bpm}$, there is nothing to guarantee the correct large $z$ behaviour (\ref{eq:Yasy}) for the $Y_{n, 1}$'s with $n \geq 1$. 
To fill this gap we will extract an extra equation from the TBA, equation (\ref{eq:Yprod}), and show that it fixes the form of the gauge transformation (\ref{eq:wTgen1})-(\ref{eq:wTgen3}). 
Finally, we will show that the final set of equations (\ref{eq:constraint31})-(\ref{eq:wT}) is self-consistent and that equation (\ref{eq:Yprod}) can be rederived from them.
\subsection*{Direct proof}
The Y function sitting at the central node can be parametrised equivalently using the gauges $\mT^{\ra}$ or $\mT^{\ua}$. 
This gives the two expressions
\bea
1 + 1/Y_{1,1} &=&
2\frac{ e^{2 \phi} \sinh( 2 \phi ) \sinh( \hB) }{  \sinh( \hB \rho_{\ra} ) } \; \frac{ { \widetilde{ \mT^{\ra\,[+1] }_{1,1} } } \;  \mT^{\ra\,[-1]}_{1,1} }{\mT^{\ra}_{1,2} }, \label{eq:oneplusY} \\
1 + Y_{1,1} &=&
-2\frac{ e^{-2 \phi} \sinh( 2 \phi ) \sinh( \hmu) }{  \sinh( \hmu \rho_{\ua} ) } \; \frac{ { \widetilde{ \mT^{\ua\,[+1] }_{1,1} } } \;  \mT^{\ua\,[-1]}_{1,1} }{   \mT^{\ua}_{2,1} }. \label{eq:oneplusY2}
\eea
Constructing the quantity
\bea
r(z) = \left(\frac{ 1 + 1/Y_{2, 2}(z) }{ 1 + Y_{1, 1}(z) }\right) = \left(\frac{ 1 + Y_{1, 1}(z) \; e^{4 \phi(z)}  }{ 1 + Y_{1, 1}(z) } \right),
\eea
 and matching the alternative expressions for $r(z)$ obtained from (\ref{eq:oneplusY}) and (\ref{eq:oneplusY2}) gives 
 the first equation of the NLIEs written in Section \ref{sec:numerics}: 
\bea\label{eq:rmatch}
r(z) = \frac{ \mT^{\ra\,[+1]}_{1,1}(z) \; { \widetilde{ \mT^{\ra\,[-1] }_{1,1} }(z) } }{\mT^{\ra\,[-1]}_{1,1}(z) \; { \widetilde{ \mT^{\ra \,[+1] }_{1,1} }(z) } } = e^{ 4\phi(z) } \frac{ \mT^{\ua\,[+1]}_{1,1}(z) \; { \widetilde{ \mT^{\ua\,[-1] }_{1,1} }(z) } }{\mT^{\ua\,[-1]}_{1,1}(z) \; { \widetilde{ \mT^{\ua\,[+1] }_{1,1} }(z) } }.
\eea
 Finding a second constraint to close the system is slightly more involved. 
 First, using (\ref{eq:oneplusY}) and (\ref{eq:oneplusY2}), we find:
\bea\label{eq:firstY}
0&=& \log{ \hY_{1,1}(z) }+4 \phi(z) -\log \left(\frac{  \sinh( \hB\rho_{\ra}(z) )/\sinh( \hB)}{  \sinh( -\hmu\rho_{\ua}(z) )/\sinh( \hmu)} \right) -  
\log \left(\frac{ { \mT }_{1,2}^{ \ra }(z)}{{ \mT }_{2,1}^{ \ua}(z)} \right) + \log\left(  \frac{{ \widetilde{ \mT^{\ra\,[+1] }_{1,1} } } \;  \mT^{\ra\,[-1]}_{1,1}  }{ { \widetilde{ \mT^{\ua\,[+1] }_{1,1} } } \;  \mT^{\ua\,[-1]}_{1,1}  }\right). \nn\\
\eea
 We can combine (\ref{eq:firstY}) with the TBA equation (\ref{eq:tba3}), simplified using the ``telescoping'' procedure discussed in Section \ref{sec:freeenergy}. 
 If the $Y_{1,n}$ and $Y_{n, 1}$ functions are parametrised using the $\mT^{\ra}$ and $\mT^{\ua}$ gauges, respectively, 
 the infinite sums in (\ref{eq:tba3}) reduce to only a few terms, giving 
\bea\label{eq:Yprod}
 \log{ \hY_{1,1}(z) }
&=& 2 \bu/T -2 \phi(z) - \log \left(\frac{ { \mT }_{1,1}^{ \ra }}{ { \mT }_{1,1}^{ \ua }} \right) \ast  a_1(z) + \log \left(\frac{ { \mT }_{1,2}^{ \ra }(z)}{{ \mT }_{2,1}^{ \ua }(z)}\right).
\eea
By using Cauchy's theorem and shifting the integration contours to infinity, 
one can write the last convolution term as
\bea\label{eq:intpassage}
\log \left(\frac{\mT^{\ra}_{1,1}}{\mT^{\ua}_{1,1}} \right) \ast a_1(z)
 &=& -\int_{-1}^{1} \frac{dv}{ 2 \pi i ( v - z ) }\; \disc\left[\log \left(\frac{ \mT^{\ra\,[+1]}_{1,1}(v) \;  \mT^{\ra\,[-1]}_{1,1} (v) }{ \mT^{\ua\,[+1]}_{1,1} (v) \;  \mT^{\ua\,[-1]}_{1,1}(v) } \right) \right] \\
 &+&\log \left(\frac{ \mT^{\ra \,[+1]}_{1,1}(z) \; \mT^{\ra\,[-1]}_{1,1}(z) }{ \mT^{\ua\,[+1]}_{1,1}(z) \;\mT^{\ua\,[-1]}_{1,1} (z) } \right). \nn
\eea
Equation (\ref{eq:Yprod}) can then be rewritten as
\bea\label{eq:pvY}
\log{ \hY_{1,1}(z) }
&=& 2 \bu/T -2 \phi(z) + \int_{-1}^{1} \frac{dv}{ 2 \pi i ( v - z ) }\; \disc\left[ \log \left(\frac{ \mT^{\ra\,[+1]}_{1,1}(v) \;  \mT^{\ra\,[-1]}_{1,1} (v) }{ \mT^{\ua\,[+1]}_{1,1} (v) \;  \mT^{\ua\,[-1]}_{1,1}(v) } \right)\right] \nn\\
&-& \log \left(\frac{ \mT^{\ra \,[+1]}_{1,1}(z) \; \mT^{\ra\,[-1]}_{1,1}(z) }{ \mT^{\ua\,[+1]}_{1,1}(z) \;\mT^{\ua\,[-1]}_{1,1} (z) } \right) + \log \left(\frac{ { \mT }_{1,2}^{ \ra }(z)}{{ \mT }_{2,1}^{ \ua }(z)} \right).
\eea
 Comparing (\ref{eq:pvY}) and (\ref{eq:firstY}) and using 
  (\ref{eq:rmatch}), we find 
\bea\label{eq:lasteqpv0}
\log \left(\frac{  \sinh^2( \hB \rho_{\ra}(z) )/\sinh^2( \hB) }{  \sinh^2(\hmu\rho_{\ua}(z) )/\sinh^2( \hmu)} \right) 
 &=& 4 \bu/T +  2  \, \int_{-1}^{1}  \disc\left[ \log \left(\frac{ \mT^{\ra\,[+1]}_{1,1}(v) \;  \mT^{\ra\,[-1]}_{1,1} (v) }{ \mT^{\ua\,[+1]}_{1,1} (v)  \;  \mT^{\ua\,[-1]}_{1,1}(v) } \right) \right]\; \frac{dv}{2 \pi i (v - z)}\nn\\
&-& \disc\left[ \log \left(\frac{ \mT^{\ra\,[+1]}_{1,1}(z) \;  \mT^{\ra\,[-1]}_{1,1} (z) }{ \mT^{\ua\,[+1]}_{1,1} (z)  \;  \mT^{\ua\,[-1]}_{1,1}(z) } \right) \right].
 \eea
Notice that the  last two terms can be nicely combined as a single principal value integral: 
\bea\label{eq:lasteqpv}
\log \left(\frac{  \sinh^2( \hB \rho_{\ra}(z) )/\sinh^2( \hB) }{  \sinh^2(\hmu\rho_{\ua}(z) )/\sinh^2( \hmu)} \right) 
 = 4 \bu/T +  2  \, \dashint_{-1}^{1}  \disc\left[ \log \left(\frac{ \mT^{\ra\,[+1]}_{1,1}(v) \;  \mT^{\ra\,[-1]}_{1,1} (v) }{ \mT^{\ua\,[+1]}_{1,1} (v)  \;  \mT^{\ua\,[-1]}_{1,1}(v) } \right) \right]\; \frac{dv}{2 \pi i (v - z)}.\nn\\
 \eea
 At this point there are different ways to proceed; perhaps the simplest strategy is to apply $$\sqrt{1-w^2}  \dashint_{-1}^{1} \frac{dv}{\pi i \sqrt{1-v^2} (v - w) }$$ to both sides of (\ref{eq:lasteqpv}).
 Because the integrand on the rhs of  (\ref{eq:lasteqpv}) is a discontinuity of square root type, it can be factorised as $\sqrt{1-z^2} g(z)$, where $g(z)$ is analytic in a neighbourhood of the interval $(-1,1)$ and therefore
 \bea\label{eq:propggg}
\frac{1}{\pi^2}\dashint_{-1}^{1} \frac{dv}{ \sqrt{1-v^2} (v - w) } \dashint_{-1}^{1} \sqrt{1-z^2} g(z) \frac{dz}{z-v} =- g(w). 
 \eea
 Thus, we find 
\bea\label{eq:eqprev}
&&- \frac{w \sqrt{1-1/w^2} }{\pi}  
\dashint_{-1}^{1} \frac{dv}{\sqrt{1-v^2} (v - w) }  \log \left(\frac{  \sinh^2( \hB \rho_{\ra}(v) )/\sinh^2( \hB) }{  \sinh^2(\hmu\rho_{\ua}(v) )/\sinh^2( \hmu)} \right)\\
&=& \disc\left[\log \left(\frac{ \mT^{\ra\,[+1]}_{1,1}(w) \;  \mT^{\ra\,[-1]}_{1,1} (w) }{ \mT^{\ua\,[+1]}_{1,1} (w) \;  \mT^{\ua\,[-1]}_{1,1}(w) } \right) \right].\nn\\
\eea
From the definitions of $\bh_{\ra}$ and $\bh_{\ua}$ given in (\ref{eq:integralh}), one can derive
\bea\label{eq:eq1app}
\text{disc}\left[\log\left(\frac{\bh_{\ra}(z)}{\bh_{\ua}(z)}\right)\right] = -\frac{z \sqrt{ 1 - 1/z^2 }}{ \pi } \; \dashint_{-1}^{1} \log\left( \frac{ \sinh(  \hmu \rho_{\ua}(v)/\sinh( \hmu )}{ \sinh(  \hB \rho_{\ra}(v) )/\sinh( \hB  ) } \right)  \frac{ dv }{ \sqrt{1 - v^2 } \; ( v - z ) },\nn\\
\eea
and the comparison with (\ref{eq:eqprev}) gives
\bea
 \disc\left[\log \left(\frac{\bh_{\ra}(z)}{\bh_{\ua}(z) }\right) \right] = \frac{1}{2} \disc\left[ \log \left(\frac{ \mT^{\ra\,[+1]}_{1,1}(z) \;  \mT^{\ra\,[-1]}_{1,1} (z) }{ \mT^{\ua\,[+1]}_{1,1} (z) \;  \mT^{\ua\,[-1]}_{1,1}(z) }\right) \right].
\eea
Together with (\ref{eq:rmatch}), this equation leads to
\bea\label{eq:closed}
\disc\left[\log\left( \frac{ \mT^{\ra \, [\pm] }_{1,1}(z)  }{  \mT^{\ua \, [\pm] }_{1,1}(z)  } \right)   + \log \left(\frac{   \bh_{\ra}(z) }{  \bh_{\ua}(z) } \right) \right]= \pm 2 \phi(z),
\eea
which can be interpreted as the statement that the combination
\bea\label{eq:term}
&&\ln \left(\frac{ \mT^{\ra  }_{1,1}(z)  }{  \mT^{\ua  }_{1,1}(z) } \right)+ \ln \left(\frac{ \bh_{\ra}^{[+1]}(z) \bh_{\ra}^{[-1]}(z) }{ \bh_{\ua}^{[+1]}(z) \bh_{\ua}^{[-1]}(z) } \right) - \phi^{[-1]}(z) + \phi^{[+1]}(z) \\
&&= \ln \left(\frac{ \wT^{\ra }_{1,1}(z)  }{  \wT^{\ua  }_{1,1}(z) } \right) - \phi^{[-1]}(z) + \phi^{[+1]}(z), \label{eq:term2}
\eea
has no cuts in the whole complex plane. Using the asymptotics specified by (\ref{eq:Cintegral}),(\ref{eq:mTasy}),(\ref{eq:asyC}), we conclude 
that the expression in (\ref{eq:term2}) is precisely zero, proving (\ref{eq:wTgen}) for the node $(1,1)$. The gauge matching at the other nodes follows by the structure of the T-system.
\subsection*{Converse proof}
Let us briefly discuss how equation (\ref{eq:pvY}) can be rederived from the statement that
\bea\label{eq:constrfinal}
\ln \left(\frac{ \wT^{\ra }_{1,1}(z)  }{  \wT^{\ua  }_{1,1}(z) } \right)  - \phi^{[-1]}(z) + \phi^{[+1]}(z) = 0,
\eea
together with the resolvent parametrisation of Section \ref{sec:resolvent}. While this may seem obvious, 
we will spell out the details in order to show that the system of nonlinear integral equations integrated in Section \ref{sec:numerics} are a consequence of (\ref{eq:constrfinal}). 
Just following backwards the steps in the derivation above and using the asymptotics $\lim_{z \rightarrow \infty} \log \left(\bh_{\ra}(z)/\bh_{\ua}(z) \right) = - \bu/T$, 
one can re-obtain equations (\ref{eq:lasteqpv})-(\ref{eq:closed}), in particular one recovers
\bea\label{eq:lasteqpv2}
\log\left(\frac{  \sinh^2( \hB \rho_{\ra}(z) )/\sinh^2( \hB) }{  \sinh^2(\hmu\rho_{\ua}(z) )/\sinh^2( \hmu)} \right) 
= 4 \bu/T +  2  \, \dashint_{-1}^{1}  \disc\left[ \log \left(\frac{ \mT^{\ra\,[+1]}_{1,1}(v) \;  \mT^{\ra\,[-1]}_{1,1} (v) }{ \mT^{\ua\,[+1]}_{1,1} (v) \;  \mT^{\ua\,[-1]}_{1,1}(v) } \right) \right]\; \frac{dv}{2 \pi i (v - z)}.\nn\\
\eea
Moreover, adding the two equations (\ref{eq:closed}) shows that the two terms in (\ref{eq:rmatch}) are equal.
We can now reverse the logic and define a function $Y_{1,1}$ satisfying 
\bea\label{eq:C20}
\left(\frac{ 1 + Y_{1, 1}(z) \; e^{4 \phi(z)}  }{ 1 + Y_{1, 1}(z) } \right) = \frac{ \mT^{\ra\,[+1]}_{1,1}(z) \; { \widetilde{ \mT^{\ra\,[-1] }_{1,1} }(z) } }{\mT^{\ra\,[-1]}_{1,1}(z) \; { \widetilde{ \mT^{\ra \,[+1] }_{1,1} }(z) } } = e^{ 4\phi(z) } \frac{ \mT^{\ua\,[+1]}_{1,1}(z) \; { \widetilde{ \mT^{\ua\,[-1] }_{1,1} }(z) } }{\mT^{\ua\,[-1]}_{1,1}(z) \; { \widetilde{ \mT^{\ua\,[+1] }_{1,1} }(z) } }.
\eea
As a consequence of (\ref{eq:C20}), $Y_{1,1}$ automatically verifies equations (\ref{eq:oneplusY}), (\ref{eq:oneplusY2}) and (\ref{eq:firstY}). Together with (\ref{eq:lasteqpv2}), this proves (\ref{eq:pvY}).
\section{Dictionary}
\label{app:dictionary}
In this Appendix, we provide a dictionary between the notation of this paper and the ones of \cite{DDVHubbard}. 
First, the conventions for the coupling constant and magnetic field are related as 
\bea
\bu^{\text{here}} = U^{\text{ref \cite{DDVHubbard}} }/4, \;\;\;\; B^{\text{here}} = H^{\text{ref \cite{DDVHubbard}} }/2
\eea
The Quantum Transfer Matrix of \cite{DDVHubbard} is characterised by three parameters: the Trotter number $\NT \in \mathbb{N}$, the inhomogeneity $u$ and the rapidity $v$. %
The Gibbs free energy at temperature $T$ can be obtained by
taking the Trotter limit 
\bea
f = -T \lim_{ \NT \rightarrow \infty } \Lambda\left( u=\frac{1}{T \NT }, \; v= 0 \right),
\eea  
where $\Lambda$ is the largest eigenvalue of the QTM.
The map between $v$ and the spectral parameter used in this paper, $s$, is\footnote{ Note that, in  this Section, we use the letter $s$ for the spectral parameter, rather than $z$ as in the rest of the paper. This is to avoid confusion with the functions $z_{\pm}(\lambda)$ defined in (\ref{eq:functionsz}) below. }
\bea\label{eq:utos}
s =  i\frac{1}{2\tan(v)} \left( \frac{\tan^2(v) -1 }{\tan^2(v) + 1} \right) \sqrt{4 \bu^2 \tan^2(v) + (1 + \tan^2(v) )^2 }.
\eea
In particular, $v = 0$ corresponds to  $s\rightarrow \infty$. \\
 The QTM eigenvalue is given by formula (15) in \cite{DDVHubbard} containing the parameters $x$, $w$, $z_{\pm}(x)$, $z_{\pm}(w)$, $h(x)$, $h(w)$ defined by
\bea
2 w &=& \log( \tan(u)),  \;\;\;\;\; 2 x = \log( \tan(v)),\\
2h(y) &=& - \arcsinh\left( \frac{\bu}{\cosh( 2 y ) } \right), \;\;\;\;\;z_{\pm}(y) = e^{2 h(y)  \pm 2 y}. \label{eq:functionsz0}
\eea
 A simple calculation shows that
\bea\label{eq:functionsz}
\hspace{-0.6cm}z_{\pm}(x) &=&  ( \tan(v)  )^{\pm1}  \, ( 1 + \tan^2(v) ) /\left(  2 \bu \, \tan(v) + \sqrt{  (1 + \tan^2(v) )^2  + 4 \bu^2 \tan^2(v)  } \right), 
\eea
 and, using (\ref{eq:utos}),
\bea
z_-(x)
&=& \x( s + i \bu )
,\;\;\;\;\;z_+(x)
=1/\x( s - i \bu ), \\ 
z_{\mp}(x) - 1/z_{\mp}(x) &=& -2 \bu \pm 2 i s, \\
 z_{\mp}(x) + 1/z_{\mp}(x) &=& -2 \phi( s \pm i \bu ).  
\eea
The Zhukovsky map $\x(z)$ is defined in (\ref{eq:Zhukovsky}), has a single short cut and satisfies $\widetilde{\x}(s) = -1/\x(s)$, and $\x(s) \sim 2 i s$ at large $s$. Using this dictionary, equation (15) in \cite{DDVHubbard} can be rewritten as
\bea\label{eq:La}
&& \hspace{-0.5cm} \Lambda(u, v)/ A_2  \left(\frac{z_+(x)}{z_-(x)} \right)^{\NT/2}\\
 &=& e^{\hmu + \hB} \; \left( A_1/A_2 + e^{\hmu - \hB} \prod_{\alpha=1}^{\NT/2} 
  \left(\frac{s -w_{\alpha} - 2 i \bu }{w_{\alpha} - s } \right) \right) \prod_{j=1}^{\NT}  \left(\frac{ 1 + 1/( \x^{[+1]}(s) z_j ) }{ -1 + \x^{[-1]}(s)/ z_j  } \right) \nn\\
  &+&  e^{\hmu - \hB} \;\left(  A_4/A_2 + e^{-\hmu + \hB} \prod_{\alpha=1}^{\NT/2} \left(\frac{ s -w_{\alpha} + 2 i \bu }{ w_{\alpha} -s } \right) \right) \, \prod_{j=1}^{\NT} 
\left(\frac{ 1 + 1/( \x^{[-1]}(s) z_j )}{ -1 + \x^{[+1]}(s)/z_j  }\right). \nn
\eea
The quantities $A_1$, $A_2$ and $A_4$ are defined in \cite{DDVHubbard} as 
\bea\label{eq:A2}
A_1/A_2 &=& \left( \frac{(1-z_-(w)z_+(x) )(1-z_+(w) z_+(x) )}{(1 + z_-(w)z_+(x) )(1+z_+(w) z_+(x) )}\right)^{\NT/2}, \\
A_4/A_2 &=& \left( \frac{(1 + z_-(w)/z_-(x) )(1 + z_+(w)/z_-(x) )}{(1 - z_-(w)/z_-(x) )(-+z_+(w)/z_-(x) )}\right)^{\NT/2},\\
A_2 &=& \left( -\cos^2(v) \cos^2(v-u)\cos^2( v + u) \right)^{\frac{\NT}{2}} \, \left(\frac{z_+(x)}{z_-(x)} \right)^{\NT/2} \\
&\times& \left( \cos^2(u) e^{2 h(w) }  \left(1 - \frac{z_-(x) }{z_-(w) } \right) \left(1 + \frac{1}{z_+(x) z_-(w) } \right) \right)^{\NT/2}.
\eea  
 Using the asymptotics $
 z_{\pm}(w) \sim ( \tan(u) )^{\pm 1} = ( \tan( \frac{1}{T \NT}) )^{\pm 1} $ for $\NT \rightarrow \infty$, it is possible to check that
\bea
\lim_{\NT \rightarrow \infty} (-1)^{\NT/2} \, A_1/A_2 &=& e^{ - (z_+(x) + 1/z_+(x))/T } =  e^{ 2 \phi^{[-1]}(s) }, \\
\lim_{\NT \rightarrow \infty} (-1)^{\NT/2} \, A_2/A_4 &=& e^{ ( z_-(x) + 1/z_-(x) )/T} =  e^{ -2 \phi^{[+1]}(s) }, 
\eea
and
\bea
&& \lim_{\NT \rightarrow \infty}\left( \cos^2(u) e^{2 h(w) }  \left(1 - \frac{z_-(x) }{z_-(w) } \right) \left(1 + \frac{1}{z_+(x) z_-(w) } \right) \right)^{\NT/2} \nn\\
 &=& e^{ -(z_-(x) - 1/z_+(x) )/(2T) - \bu/T  } = e^{ \frac{1}{2} (\phi^{[+1]}(s) -\phi^{[-1]}(s) )}.
\eea
Therefore, we can rewrite (\ref{eq:La}) in the limit of large, even $\NT$ as
\bea\label{eq:La2}
&&\Lambda(u, v) \; e^{- \frac{1}{2} (\phi^{[+1]}(s) -\phi^{[-1]}(s) )} \; \left( \cos^2(v) \cos^2(v-u)\cos^2(v+u) \right)^{-\NT/2} \nn\\
&\sim& e^{\hmu + \hB} \left(  e^{ 2 \phi^{[-1]} } + e^{\hmu - \hB} \prod_{\alpha=1}^{\NT/2}\left(\frac{s -w_{\alpha} - 2 i \bu }{ s - w_{\alpha}} \right) \right) \prod_{j=1}^{\NT} \left(\frac{ 1 + 1/(\x^{[+1]}(s) z_j ) }{ 1 - \x^{[-1]}(s)/ z_j  } \right) \nn\\
&+& e^{\hmu - \hB} \left(  e^{ -2 \phi^{[+1]} } + e^{-\hmu + \hB} \prod_{\alpha=1}^{\NT/2} \left(\frac{ s -w_{\alpha} + 2 i \bu }{ s - w_{\alpha}} \right) \right) \, \prod_{j=1}^{\NT} \left(\frac{ 1 + 1/(\x^{[-1]}(s) z_j )}{ 1 - \x^{[+1]}(s)/z_j  }\right).
\eea
The factor $\left( \cos^2(v) \cos^2(v-u)\cos^2(v+u) \right)^{-\NT/2}$ is simply a  normalization (divergent in the $\NT \rightarrow \infty$ limit for $v \neq 0$, and converging to one for $v=0$). We expect that, once this term is factored out, the limit $\bar{N} \rightarrow \infty$ can be taken in the rhs of (\ref{eq:La2})  without first setting $v=0$, giving a well-defined function of $v$ (or, equivalently, of $s$).\\
Considering the identifications discussed in Section \ref{sec:QTM},
\bea
\bF_{\p \m}(s) &=& e^{-\frac{i}{2} ( \hB- \hmu ) s/\bu } \lim_{ \NT \rightarrow \infty } \prod_{\alpha=1}^{\NT/2} \left( s - w_{\alpha} \right), \\
 \bP^{\ra}_{\bpm} &=& e^{ \mp \frac{i}{2} \hB s/\bu } \lim_{\NT \rightarrow \infty }  C_{\ra} \; \prod_{j=1}^{\NT} \left( 1 + 1/(\x(s) z_j ) \right),\\
 z_j &=& \x(s_j ),
\eea
we find a nice agreement between (\ref{eq:La2}) and (\ref{eq:idenprecise}), leading to
\bea
\lim_{\NT \rightarrow \infty} \Lambda(u, v) \; \left( \cos^2(v) \cos^2(v-u)\cos^2(v+u) \right)^{-\NT/2} = e^{ \hmu + \frac{1}{2} (\phi^{[+1]}(s) -\phi^{[-1]}(s) )} \; \wT^{\ra}_{1,1}(s).\nn\\
\eea 

\bibliography{biblio3} 
\end{document}